\documentclass[twocolumn,aps,pra,superscriptaddress,10pt]{revtex4-2} 
\usepackage{graphicx}  
\usepackage{dcolumn}   
\usepackage{bm}        
\usepackage{amssymb}  
\usepackage{amsmath}   
\UseRawInputEncoding
\usepackage{hyperref}
\usepackage[version=4]{mhchem}
\usepackage{siunitx}
\usepackage{natbib}
\usepackage{braket}
\usepackage{appendix}
\usepackage{soul}
\renewcommand{\vec}[1]{\boldsymbol{#1}}

\begin{document}

\title{Exact kinetic propagators for coherent state complex Langevin simulations} 
\author{Thomas G. Kiely}
\thanks{Equal contribution}
\affiliation{Kavli Institute of Theoretical Physics, University of California, Santa Barbara, 93106, California, USA}

\author{Ethan C.~McGarrigle} 
\thanks{Equal contribution}
\affiliation{Department of Chemical Engineering, University of California, Santa Barbara, 93106, California, USA}
 
\author{Glenn H.~Fredrickson}
\email{ghf@ucsb.edu}
\affiliation{Department of Chemical Engineering, University of California, Santa Barbara, 93106, California, USA}
\affiliation{Materials Research Laboratory, University of California, Santa Barbara, 93106, California, USA}
\affiliation{Materials Department, University of California, Santa Barbara, 93106, California, USA}

\date{\today}

\begin{abstract}
We introduce and benchmark an improved algorithm for complex Langevin simulations of bosonic-coherent state path integrals. Our approach utilizes a Strang splitting of the imaginary-time propagator rather than the conventional linear-order Taylor expansion, allowing us to construct an action that incorporates
higher-order terms at negligible computational cost. The resulting algorithm enjoys guaranteed linear stability independent of the imaginary-time discretization, enabling more resource-efficient simulations. We demonstrate this improved performance
for single-species bosons and for two-component bosons with Rashba spin-orbit coupling. 
\end{abstract}

\maketitle
\section{Introduction}
Numerical path integral methods provide an exact approach for solving equilibrium quantum many-body problems, capturing the nontrivial interplay between quantum and thermal fluctuations
~\cite{shumwayPathIntegralMonte2000, shenFiniteTemperatureAuxiliary2020, rubensteinFinitetemperatureAuxiliaryfieldQuantum2012, shenStableRecursiveAuxiliary2023}. These methods have had remarkable success at modeling interacting ensembles of bosons, with notable examples being
superfluid $^4{\rm He}$~\cite{ceperleyPathIntegralsTheory1995, ceperleyPathintegralComputationLowtemperature1986, ceperleyPathintegralSimulationSuperfluid1989, herdmanEntanglementAreaLaw2017, herdmanQuantumMonteCarlo2014} and ultracold atom Bose-Einstein condensates (BECs) coupled to artificial gauge fields~\cite{hayataComplexLangevinSimulation2015,mcgarrigleEmergenceSpinMicroemulsion2023,attanasioThermodynamicsSpinorbitcoupledBosons2020, fredricksonFieldTheoreticSimulationsSoft2023}. 
Path integral methods rely on a well-known mapping between a $d$-dimensional quantum partition function, $\mathcal{Z}=\text{Tr}[\exp(-\beta\hat H)]$, and a $(d+1)$-dimensional classical path integral~\cite{feynmanSpaceTimeApproachNonRelativistic1948}. While the mapping is unambiguous in the continuum limit, numerical methods require one to discretize the path integral in the imaginary-time coordinate $\tau \in (0,\beta)$. There are many choices of how one constructs the discretized path integral; while all are equivalent in the continuum limit, this choice can drastically change the accuracy and stability of numerical methods. ``Higher-order" discretized path integrals have been shown in various cases to improve numerical simulations~\cite{chinGradientSymplecticAlgorithms2002}, but these approaches require additional numerical resources.
In this work we provide a general means of constructing numerically-stable coherent-state path integrals from bosonic quantum Hamiltonians without demanding any additional resources. Our method is complementary to higher-order constructions: It can seamlessly be integrated into these approaches, but it provides improved stability during complex Langevin sampling even at linear order. We benchmark our method 
in two experimentally-relevant scenarios: the single-component Bose gas and the two-component Bose gas with Rasbha spin-orbit coupling.

Neutral atom arrays 
and Bose-Einstein condensates
form a rich starting point for realizing various 
exotic states of matter, such as topological phases \cite{semeghiniProbingTopologicalSpin2021, cooperTopologicalBandsUltracold2019, liTopologicalStatesLadderlike2013, valdes-curielTopologicalFeaturesLattice2021}, supersolids \cite{liStripePhaseSupersolid2017, blandTwoDimensionalSupersolidFormation2022, norciaTwodimensionalSupersolidityDipolar2021}, and quantum hall analog states \cite{mukherjeeCrystallizationBosonicQuantum2022}. These states are often theoretically understood in terms of simplified limits and toy models, such as the Laughlin state and the Toric code. Experimental realizations, however, rarely look so simple.
Mean-field methods~\cite{pitaevskiiBoseEinsteinCondensationSuperfluidity2016}, which have had great success at modeling weakly-interacting BECs, introduce uncontrolled approximations when applied to these interacting and strongly-correlated systems. 

Numerical path integral approaches, which treat quantum and thermal fluctuations on the same footing, provide a path towards numerically-exact and experimentally-realistic modeling of bosonic ensembles. One such technique, path integral Monte Carlo (PIMC) \cite{polletRecentDevelopmentsQuantum2012}, uses Monte Carlo sampling over the coordinate basis to compute averages over the full path integral, requiring explicit exchange permutation sampling to enforce Bose statistics. This method provided some of the earliest successes in numerical path integral techniques~\cite{ceperleyPathIntegralsTheory1995, ceperleyPathintegralComputationLowtemperature1986, ceperleyPathintegralSimulationSuperfluid1989} and has been used to study a wide range of strongly-interacting models in the continuum ~\cite{herdmanEntanglementAreaLaw2017, herdmanQuantumMonteCarlo2014} and lattice \cite{mahmudFinitetemperatureStudyBosons2011, capogrosso-sansoneMonteCarloStudy2008} settings. The PIMC method, however, suffers from the sign problem and has generally been constrained to small system sizes of $\lesssim10^3$ particles.

Coherent state (CS) field theories with complex Langevin (CL) sampling \cite{delaneyNumericalSimulationFiniteTemperature2020, fredricksonFieldTheoreticSimulationsSoft2023, heinenComplexLangevinApproach2022} have emerged as a robust way to interrogate large bosonic assemblies with moderate interactions, regardless of whether a sign problem is present. Ultracold boson systems with artificial gauge fields \cite{goldmanLightinducedGaugeFields2014} often introduce an explicit sign problem and constitute a natural application for CSCL. Heretofore, CSCL has enabled finite-temperature investigations of rotating BECs \cite{hayataComplexLangevinSimulation2015} and spin-orbit coupled BECs \cite{mcgarrigleEmergenceSpinMicroemulsion2023,attanasioThermodynamicsSpinorbitcoupledBosons2020}. In the case of isotropic two-dimensional ``Rashba'' spin-orbit coupling \cite{goldmanLightinducedGaugeFields2014, manchonNewPerspectivesRashba2015, galitskiSpinOrbitCoupling2013}, the massive single-particle degeneracy greatly enhances the role of fluctuations, and CSCL has unveiled rich finite temperature behavior with a prediction of a quantum microemulsion analog \cite{mcgarrigleEmergenceSpinMicroemulsion2023, mcgarrigleThermodynamicStabilitySpin2024}. However, the applicability of CSCL is constrained by the memory cost of storing the coherent state fields in the resolved $(d+1)$ space-imaginary time dimensions, where the standard first-order accurate theory requires fine imaginary time discretizations to ensure convergent contributions to and unbiased sampling of the partition function \cite{delaneyNumericalSimulationFiniteTemperature2020}.  

In this work, we introduce a superior algorithm for performing CSCL simulations of continuum Bose fluids.
Our method consists of a judicious second-order splitting of the imaginary-time propagator and utilizes the property that exponentials of quadratic operators, $\exp(\hat O)$, map coherent states to coherent states.
The resulting classical action incorporates many higher-order terms at a negligible computational cost and enjoys guaranteed linear numerical stability, independent of the imaginary-time discretization. We demonstrate the power and flexibility of this approach by applying it to a one-component Bose gas and to a two-component, two-dimensional Bose gas with Rasbha spin-orbit coupling.

\section{Preliminaries}
Here, we utilize the complex Langevin method for simulating equilibrium coherent-state path integrals. Coherent state wavefunctions \cite{negeleQuantumManyparticleSystems1988}, which take the form  $|\phi\rangle=\exp\left(\sum_\alpha \phi(\alpha) \hat \psi^\dagger(\alpha)\right)|0\rangle$ where $\hat \psi^\dagger(\alpha)$ is the boson creation operator in state $\alpha$ \cite{fetterQuantumTheoryManyParticle2012} and $\phi(\alpha)$ is a complex scalar field, constitute an overcomplete basis and admit a resolution of the identity. The equilibrium CS path integral is constructed by inserting $N_\tau$ such identity operators
into the 
partition function, $\mathcal{Z}={\rm Tr}[e^{-\beta\hat H}]$, where $\hat H$ is a bosonic Hamiltonian and $\beta=1/k_BT$. The partition function may then be rewritten
as a path integral over the complex-conjugate fields $\phi_j$ and $\phi_j^*$,
\begin{equation}\label{eq:Z}
\begin{split}
    \mathcal{Z}&=\int \mathcal{D}(\phi^*,\phi)e^{-\sum_{j=0}^{N_\tau-1}\sum_\alpha\phi_j^*(\alpha)\phi_j(\alpha)}\\
    &\qquad\qquad\qquad\times\prod_{j=0}^{N_\tau-1}\langle\phi_j|e^{-\Delta\hat H}|\phi_{j-1}\rangle,
\end{split}
\end{equation}
weighted by matrix elements
of the imaginary-time propagator. In the above we have defined an imaginary timestep $\Delta=\beta/N_\tau$, the periodic index $j$ for distinct resolutions of the identity (i.e. $\phi_{N_\tau}\equiv\phi_0$).

Numerical \cite{fredricksonFieldTheoreticSimulationsSoft2023} and analytic \cite{sachdevQuantumPhaseTransitions2001, negeleQuantumManyparticleSystems1988} approaches to studying the equilibrium path integral require a means of optimizing or 
sampling the field variables $\phi_j(\alpha)$. A systematic approach for doing this is to rewrite the partition function as $\mathcal{Z}=\int \mathcal{D}(\phi^*,\phi)\ e^{-S[\phi^*,\phi]}$ in terms of an action functional $S[\phi^*,\phi]$. 
Given $S[\phi^*,\phi]$, both mean-field solutions and exact equilibrium thermodynamics can be accessed by the following descent scheme \cite{manCoherentStatesFormulation2014, delaneyNumericalSimulationFiniteTemperature2020}:
\begin{eqnarray}\label{eq:phi_eom}
    \frac{\partial \phi_{j}(\alpha)}{\partial t} &=& - \frac{\delta S[\phi, \phi^*]}{\delta \phi^*_{j} (\alpha)} + \eta_{j}(\alpha) , \\
    \label{eq:phistar_eom}
    \frac{\partial \phi^*_{j}(\alpha)}{\partial t} &=& - \frac{\delta S[\phi, \phi^*]}{\delta \phi_{j} (\alpha)} + \eta^*_{j}(\alpha) ,
\end{eqnarray}
where $\eta$ and $\eta^*$ denote complex-conjugate noise sources with zero mean and unit variance \cite{fredricksonFieldTheoreticSimulationsSoft2023}. Omitting the noise constitutes an imaginary-time relaxation algorithm to find $\tau$ (or $j$)-independent saddle point solutions. Including the noise corresponds to a stochastic complex Langevin dynamics that provides unbiased sampling of the coherent state theory. State of the art algorithms for integrating Eqs.~(\ref{eq:phi_eom}) and (\ref{eq:phistar_eom}) invoke  exponential time differencing (ETD) schemes that can be used to sample the stationary distribution $\propto e^{-S}$ with accuracy up to $\mathcal{O}(\Delta t^2)$, where $\Delta t$ is the Langevin timestep. 

Constructing $S[\phi^*,\phi]$ from Eq.~(\ref{eq:Z}) requires one to evaluate the matrix elements $\langle\phi_j|e^{-\Delta\hat H}|\phi_{j-1}\rangle$, which in most cases can only be done approximately.
The textbook approach \cite{negeleQuantumManyparticleSystems1988, fredricksonFieldTheoreticSimulationsSoft2023} is to Taylor expand to linear order in $\Delta$. The expectation value $\langle\phi_j|\hat H|\phi_{j-1}\rangle=\langle\phi_j|\phi_{j-1}\rangle H[\phi_j^*,\phi_{j-1}]$ is trivial after normal-ordering the bosonic field operators. 
Resumming the result, one finds
\begin{equation}
    \langle\phi_j|e^{-\Delta\hat H}|\phi_{j-1}\rangle=\langle\phi_j|\phi_{j-1}\rangle e^{-\Delta H[\phi_j^*,\phi_{j-1}]}+\mathcal{O}(\Delta^2).
\end{equation}
The action functional is then
\begin{equation}\label{eq:csaction_old}
    \begin{split}
        S[\phi^*,\phi]&=\sum_{j=0}^{N_\tau-1}\sum_\alpha~\phi^*_j(\alpha)(\phi_{j}(\alpha)-\phi_{j-1}(\alpha)) \\ &+\Delta\sum_{j=0}^{N_\tau-1}H[\phi^*_j,\phi_{j-1}] +\mathcal{O} (\Delta).
    \end{split}
\end{equation}
where the overlap between coherent states has been inserted in the form $\langle \phi_j |\phi_{j-1}\rangle = \exp [\sum_\alpha \phi_j^* (\alpha ) \phi_{j-1} (\alpha )]$. We note that the familiar (real-space) form of Eq.~(\ref{eq:csaction_old}) is obtained by replacing $\alpha\to\vec r$ and $\sum_\alpha\to\int_{V}d^dr$.


\section{Improved algorithm}
The innovation in this work is the use of a more sophisticated approximation for evaluating the imaginary-time propagator. The resulting algorithm is still strictly linear-order in $\Delta$, but incorporates higher-order terms and exhibits favorable numerical stability properties for Langevin sampling.
Our strategy starts from a second-order Strang splitting of the imaginary-time propagator, $e^{-\Delta\hat H}=e^{-\Delta\hat H_0/2}e^{-\Delta\hat H_1}e^{-\Delta\hat H_0/2}+\mathcal{O}(\Delta^3)$, where we have separated the Hamiltonian into two terms $\hat H=\hat H_0+\hat H_1$. While this decomposition is generic, in the present case we require that (1) $\hat H_0$ is quadratic in the bosonic creation/annihilation operators and (2) the minimum eigenvalue of $\hat H_0$ is zero. Provided the spectrum is lower-bounded, the latter condition can always be fixed 
by adding and subtracting quadratic terms from $\hat H_0$ and $\hat H_1$, respectively.
Henceforth we will suggestively refer to $\hat H_0$ as the {\it kinetic} term in the Hamiltonian and $\hat H_1$ as the {\it interaction} term.

Provided that $\hat H_0$ satisfies condition (1), we may 
{\it exactly} rewrite $\langle\phi_j|e^{-\Delta\hat H_0/2}e^{-\Delta\hat H_1}e^{-\Delta\hat H_0/2}|\phi_{j-1}\rangle$ as the expectation value of the interacting propagator, $\langle\phi_j^\prime|e^{-\Delta\hat H_1}|\phi_{j-1}^\prime\rangle$,
with respect to transformed wavefunctions $|\phi^\prime_{j-1}\rangle=e^{-\Delta\hat H_0/2}|\phi_{j-1}\rangle$. Importantly, the states $|\phi^\prime_j\rangle$ are still coherent states. Thus this approach requires minimal alterations to existing CSCL implementations.

It is most straightforward to construct the transformed wavefunctions in the eigenbasis of $\hat H_0$. In second-quantized notation, we define $\hat H_0=\sum_\lambda\epsilon_\lambda \hat \psi^\dagger(\lambda) \hat \psi(\lambda)$. One may freely rewrite the coherent state wavefunction in this basis, $\exp(\sum_\lambda\phi(\lambda)\hat \psi^\dagger(\lambda))|0\rangle$, using the unitary matrix $U^\dagger_{\alpha,\lambda}=\langle 0|\hat \psi(\alpha)\hat \psi^\dagger(\lambda)|0\rangle$~\cite{si}. The transformed wavefunction is then given by~\cite{si,gerryIntroductoryQuantumOptics2004,blasiakCombinatoricsBosonNormal2007}
\begin{eqnarray}
    |\phi^\prime\rangle&=&\exp\left(\sum_\lambda e^{-\Delta\epsilon_\lambda/2}\phi(\lambda)\hat \psi^\dagger(\lambda) \right)|0\rangle \\
    \langle\phi^\prime|&=&\langle 0|\exp\left(\sum_\lambda e^{-\Delta\epsilon_\lambda/2}\phi^*(\lambda)\hat \psi(\lambda) \right)
\end{eqnarray}
Henceforth we will refer to the transformed fields themselves, which are defined in the diagonal basis as $\phi^\prime(\lambda)=e^{-\Delta\epsilon_\lambda/2}\phi(\lambda)$ but may be subsequently transformed into other bases throughout the calculation.

We evaluate the resulting matrix element, $\langle\phi_j^\prime|e^{-\Delta\hat H_1}|\phi_{j-1}^\prime\rangle$, using the conventional linear-order Taylor expansion, but now with respect to the transformed CS wavefunctions:
\begin{equation}
    \langle\phi_j^\prime|e^{-\Delta\hat H_1}|\phi_{j-1}^\prime\rangle=\langle\phi_j^\prime|\phi_{j-1}^\prime\rangle e^{-\Delta H_1[(\phi^*_j,\phi_{j-1})^\prime]}+\mathcal{O}(\Delta^2)
\end{equation}
The resulting action functional,
\begin{equation}
    \begin{split}
        S[\phi^*,\phi]&=\sum_{j=0}^{N_\tau-1}\sum_\lambda~\phi^*_j(\lambda)[\phi_{j}(\lambda)-e^{-\Delta\epsilon_\lambda}\phi_{j-1}(\lambda)] \\ &+\Delta\sum_{j=0}^{N_\tau-1}H_1[(\phi^*_j,\phi_{j-1})^\prime],
    \end{split}
    \label{eq: action_new}
\end{equation}
is accurate to linear order in $\Delta$, like Eq.~(\ref{eq:csaction_old}), but incorporates many higher-order terms without introducing significant algorithmic costs. For example, Eq.~(\ref{eq: action_new}) is exact to all orders in $\Delta$ if $\hat H_1=0${, with an example calculation shown in \cite{si} where observables are independent of $\Delta$.} We emphasize that $H_1$ is a functional of the {\it transformed} fields, so functional derivatives of this term in the action (as appear in Eqs.~(\ref{eq:phi_eom}) and~(\ref{eq:phistar_eom})) will include a Jacobian factor, with an example detailed in \cite{si}.

This exact treatment of the kinetic term has important implications for the stability of the Langevin dynamics.
The quadratic term in the action contributes a term $\sum_lA_{jl}\phi_l$ to the equation of motion for $\phi_j$ (see Eq.~(\ref{eq:phi_eom}) and similar for $\phi_j\leftrightarrow\phi_j^*$). Linear stability of the algorithm is determined by the signs of the eigenvalues of $A$. For the improved action functional in Eq.~(\ref{eq: action_new}), the algorithm is linearly stable if ${\rm Re}[1\pm e^{-\Delta\epsilon_\lambda}]\geq0~\forall~\lambda$. Thus the algorithm enjoys absolute linear stability if $\hat H_0$ has a positive (semi-)definite spectrum (see condition (2) above). By contrast, the action functional in Eq.~(\ref{eq:csaction_old}) requires ${\rm Re}[1\pm(1 - \Delta\epsilon_\lambda)]\geq 0~\forall~\lambda$, and hence is only stable for sufficiently small $\Delta=\beta/N_\tau$. Stability in the original approach then requires more $N_{\tau}$ discretization, making the method more resource-intensive as the required memory to store the CS fields and computation time per CL step increases. This stability constraint is exacerbated when using a fine spatial grid to model translationally-invariant systems, as this requires the incorporation of large-momentum (high-energy) modes.


In order to compute observables, $\langle\hat O(\vec r)\rangle$, one introduces an additional term to the Hamiltonian, $\hat H_J=-\beta^{-1}\int d\vec r J(\vec r)\hat O(\vec r)$. This endows the partition function $\mathcal{Z}[J]$ with a dependence on the source fields, $J(\vec r)$. The expectation value is obtained as $\langle\hat O(\vec r)\rangle=\delta\ln\mathcal{Z}[J]/\delta J(\vec r)\bigr|_{J(\vec r) \to 0}$. In practical CSCL simulation, this procedure yields an analytic expression for a functional $O[\phi,\phi^*]$ whose average with respect to the CS path integral is $\langle\hat O(\vec r)\rangle$~\cite{fredricksonFieldTheoreticSimulationsSoft2023}. As we show in Ref.~\cite{si}, the functionals for computing observable expectation values in this manner retain the same form as in the primitive method, but with respect to the transformed fields rather than the bare fields: $\tilde O[\phi,\phi^*]=O[(\phi,\phi^*)^\prime]$.



\section{Results}
To benchmark the performance of Eq.~(\ref{eq: action_new}), we first consider a two-dimensional (2D) interacting Bose gas.
The Hamiltonian may be written as $\hat H_0+\hat H_1$ with
\begin{eqnarray}
    \hat H_0&=&\int d\vec r\ \hat \psi^\dagger (\vec r)\left(-\frac{\hbar^2}{2m}\vec\nabla^2\right) \hat \psi(\vec r) \\
    \hat H_1&=& \frac{1}{2}\int d\vec r\hspace{-0.1cm} \int d\vec r' \hat \psi^\dagger(\vec r)\hat \psi^\dagger(\vec r') u(\vec r - \vec r') \hat \psi (\vec r) \hat \psi (\vec r'),~~~~~ 
\end{eqnarray}
\noindent{where} in our grand canonical formulation, we include the chemical potential contribution in $\hat{H}_{1} \to \hat{H}_{1} -\mu\hat N$.
Here $\hat H_0$ is diagonalized with a Fourier transform, so our diagonal basis is $\lambda\equiv\vec k$ with eigenvalues 
$\epsilon(\vec k)=\hbar^2k^2/2m$. 
Throughout this work we consider a delta function pseudopotential $u(\vec r - \vec r') = g \delta(\vec r - \vec r')$, a good approximation for s-wave collisions between ultracold atoms, where the strength $g$ is directly related to the s-wave scattering length $a_{s}$ via {$g_{\rm 3D} = 4\pi \hbar^2 a_{s}/m$} \cite{pitaevskiiBoseEinsteinCondensationSuperfluidity2016}. We emphasize that our technique applies equally well to long-range interactions.
We present the results in dimensionless form, rescaled by the healing length $\ell = (\hbar^2/2m\mu)^{1/2}$ and chemical potential $\mu$ such that $\bar{\beta}=\beta \mu$ constitutes a dimensionless inverse temperature. {For both examples below, we use two-dimensional simulations that assume a quasi-2D environment with axial length scale $\ell_{z} = \sqrt{\hbar/m \omega_{z}}$ with $\omega_{z}$ an axial harmonic trapping frequency. As such, an effective two-dimensional repulsive coupling constant is determined by $g_{\rm 2D} = g_{\rm 3D} / \sqrt{2\pi} \ell_{z}$, leading to a dimensionless constant $\bar{g}_{\rm 2D} = 2mg_{\rm 2D}/\hbar^2$.} 
We perform simulations in periodic cells with rescaled sizes $\bar{L}_{\nu} = L_{\nu}/\ell$ in each direction $\nu = x,y$. 

\begin{figure}[h]
    \centering
\includegraphics[width=0.95\linewidth]{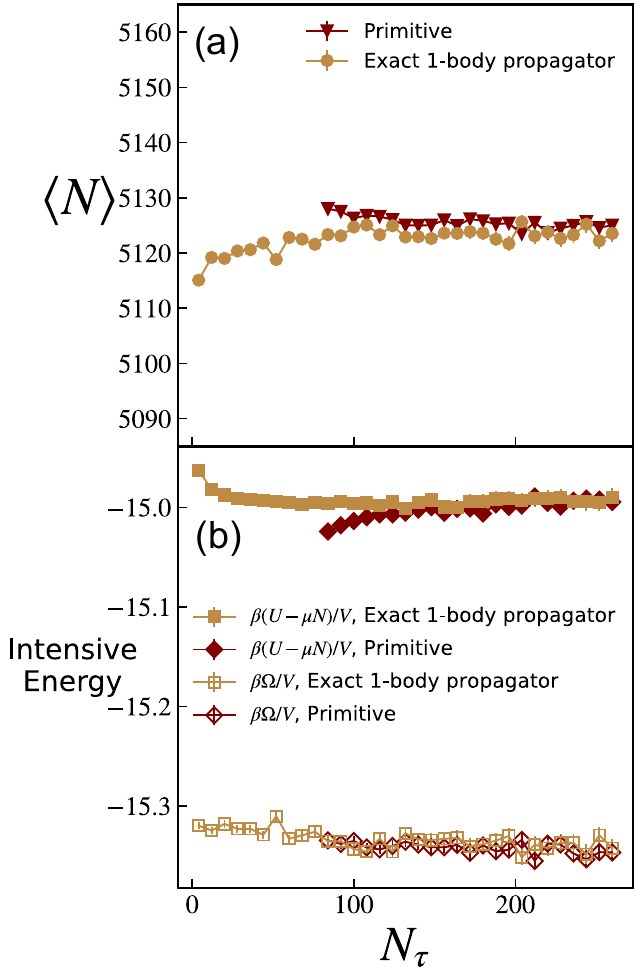}
\caption{Demonstration of $N_{\tau}$ convergence on a single component, two-dimensional Bose gas with contact interactions {$\bar{g}_{\rm 2D} = 0.0165$} at $\bar{T} = \bar{\beta}^{-1} = 2.0$. Convergence plots for the a) particle number as well as the b) intensive canonical internal energy (filled markers) and grand free energy (open markers), showing the standard first order ``primitive'' treatment with maroon diamonds and the exact one-body propagator method introduced in this work with gold squares. At lower $N_{\tau}$, the missing primitive data is due to numerical instabilities where data cannot be collected. Simulations used the ETD algorithm with $\Delta t = 0.005$ and were conducted in a square cell with side length $\bar{L} = 9.19$ with $N_{x} = 36$ plane waves in each direction. }
    \label{fig: 1-component_BF}
\end{figure}

In Fig.~(\ref{fig: 1-component_BF}) we show the results of a CSCL simulation of the 2D Bose gas using our new approach (in gold) as well as the ``primitive" first-order approach (Eq.~(\ref{eq:csaction_old}); in maroon) for various $N_\tau$ and $\bar{\beta} = 0.5$, {$\bar{g}_{\rm 2D} = 0.0165$}. We find that the primitive approach requires $N_\tau>72$ to achieve numerical stability. Our new approach, by contrast, is numerically stable down to the smallest $N_\tau$ sampled here ($N_{\tau}=4$). The two methods show excellent agreement at large $N_\tau$, with the new approach maintaining accuracy within 0.2 \% for the particle number out to very coarse imaginary time resolutions (small $N_\tau$). 
The small numerical difference in canonical internal energy $U - \mu N$ and the grand free energy $\Omega$ in Fig.\ (\ref{fig: 1-component_BF}) highlights the low entropy per particle in the superfluid phase with quasi-long range order.  

We now consider a nontrivial example with an explicit sign problem:
a two-component, two-dimensional Bose fluid with Rashba spin-orbit coupling. The Hamiltonian is
\begin{eqnarray}
    \hat H_0&=&\frac{1}{2m}\sum_{\alpha\gamma}\int d\vec r\ \hat \psi^\dagger_{\alpha}(\vec r) \left(\delta_{\alpha\gamma}i\hbar\vec\nabla-\vec{\mathcal A}_{\alpha\gamma}\right)^2 \hat \psi_{\gamma}(\vec r)~~~~ \\
    \hat H_1&=&\sum_{\alpha\gamma}\frac{g_{\alpha\gamma}}{2}\int d\vec r\hspace{-0.05cm} ~\hat \psi^\dagger_{\alpha} (\vec r)\hat \psi^\dagger_{\gamma} (\vec r) \hat \psi_{\gamma} (\vec r) \hat \psi_{\alpha} (\vec r)~~~
\end{eqnarray}
where $\alpha,\gamma$ are spin indices and {$g_{\alpha\gamma}=g_{\rm 2D}(\delta_{\alpha \gamma} + \eta(1 - \delta_{\alpha \gamma}))$} is a symmetric spin-dependent contact interaction strength, where $\eta > 1$  $(<1)$ denotes immiscible (miscible) conditions between the two components.
The vector potential is given by $\vec{\mathcal{A}}_{\alpha\gamma}=\hbar\kappa(\sigma^x_{\alpha\gamma}\hat x+\sigma^y_{\alpha\gamma}\hat y)$ where $\sigma^a_{\alpha\gamma}$ are matrix elements of the Pauli vectors in spin space ($a = x, y, z)$.
The kinetic term $\hat H_0$ has two bands with eigenvalues
$\epsilon_\pm(\vec k)=\hbar^2(k^2\pm2\kappa|\vec k|)/2m$. 
Following Ref.\ \cite{mcgarrigleEmergenceSpinMicroemulsion2023},
we report our results in a dimensionless form defined by the characteristic energy $\mu_{\text{eff}} = \mu - \hbar^2\kappa^2/m$ and healing length $\ell = (\hbar^2/2m\mu_{\text{eff}})^{-1/2}$. This procedure yields 
a dimensionless SOC strength $\bar{\kappa} = \kappa \ell$. See Ref.~\cite{si} for more details on our implementation.

Our new approach requires us to work in the eigenbasis of $\hat H_0$ (see Ref.~\cite{si} for details for building the coherent state path integral and corresponding CL algorithm). We furthermore must shift $\hat H_0$ such that it is positive-definite; this is achieved by an unbiased shift $\hat{H}_{0} \to \hat{H}_{0} + \mu'\hat{N}$, $\hat{H}_{1} \to \hat{H}_{1} - \mu'\hat{N}$ so that $\hat{H} = \hat{H}_{0} + \hat{H}_{1}$ remains unchanged. As such, we set $\mu' = \min_{\vec k} \epsilon_{-}(\vec k) = \bar{\kappa}^2$.  
\begin{figure}[h]
    \centering %
    \includegraphics[scale=0.72]{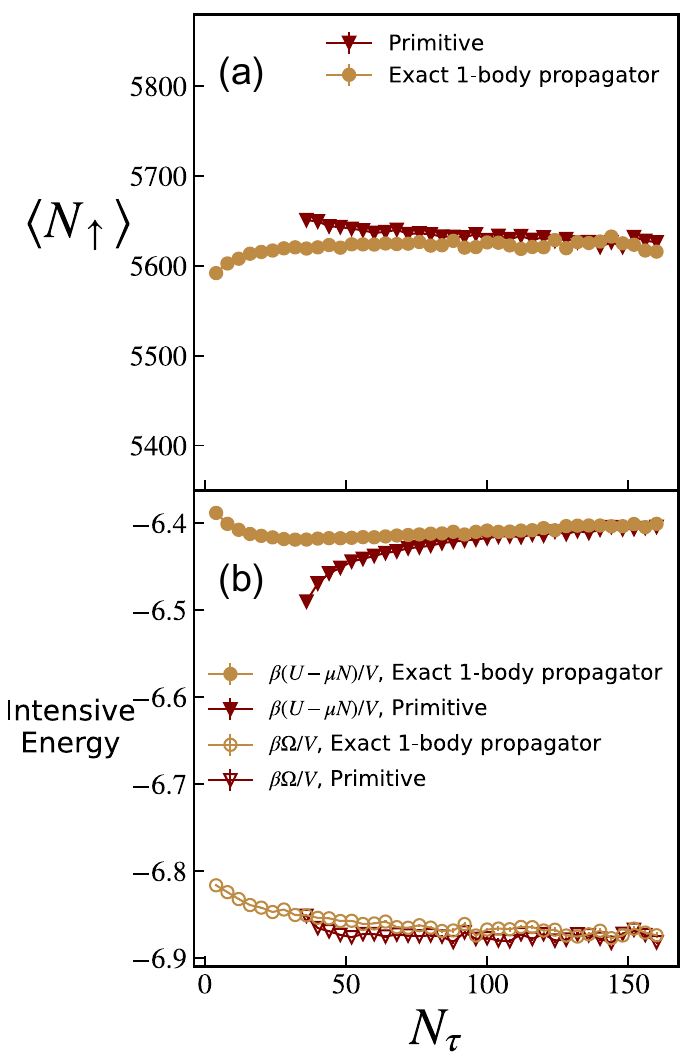} 
    \caption{Demonstration of $N_{\tau}$ convergence on a two-component, two-dimensional Rashba spin-orbit coupled Bose gas in the stripe phase with contact interactions {$\bar{g}_{2D} = 0.1$} at $\bar{T} = \bar{\beta}^{-1} = 1.0$ in immiscible conditions $(\eta = 1.1)$. Convergence plots for the a) particle number as well as the b) intensive canonical internal energy (filled markers) and grand free energy (open markers), showing the standard first order ``primitive'' treatment with maroon triangles and the exact one-body propagator method with gold circles. At lower $N_{\tau}$, we note missing primitive data due to numerical instabilities where data cannot be collected. Simulations used the ETD algorithm with $\Delta t = 0.01$ and were conducted in a square cell with side length $\bar{L} = 10\pi$ with $N_{x} = 56$ plane waves in each direction. }
    \label{fig: SOC_demo}
\end{figure}

In Fig.\ (\ref{fig: SOC_demo}) we show the results of CSCL simulations of the 2D Rashba spin-orbit-coupled Bose gas, contrasting our new approach (in gold) with the primitive propagator (in maroon) for various $N_\tau$. Here the system is in a superfluid stripe phase with smectic spin order~\cite{mcgarrigleEmergenceSpinMicroemulsion2023}. We find that the primitive approach becomes unstable for $N_\tau\leq 35$, while again the new approach remains stable down to remarkably coarse imaginary-time grids ($N_\tau = 4$). We observe good convergence for the new method as $N_{\tau}$ increases, noting that the observable estimates at small $N_{\tau}$ show less than $1\%$ deviation from the large $N_{\tau}$ value. In Figure\ (\ref{fig: SOC_demo}b), the primitive estimate of the internal energy converges more slowly than our new approach.

\section{Discussion}
The demonstrations in Figures~(\ref{fig: 1-component_BF}) and (\ref{fig: SOC_demo}) represent a significant improvement in method efficiency. 
With the simple change proposed here, the same thermodynamic results are obtained with significantly reduced computational cost. The reduction in required $N_{\tau}$ leads to appreciable savings in both memory and in computation time per CL step. In bypassing the linear stability requirement from the primitive method, the new propagator treatment permits high-resolution simulations of bosonic matter down to very low temperatures. 

The case of Rashba SOC presents an example where diagonalizing during the path integral procedure leads to efficiency improvements when conducting Langevin sampling  downstream. Since the Langvin SOC drift terms are diagonal in the dressed state basis, ETD algorithms are efficient and integrate the SOC terms to all orders in $\Delta t$. The improved accuracy and stability of this approach could enable efficient simulations of strongly spin-orbit coupled bosons ($\kappa \gg 1$), where rich phenomena are speculated \cite{gopalakrishnanQuantumQuasicrystalsSpinOrbitCoupled2013, wilsonMeronGroundState2013}.

As noted earlier, resolution in the imaginary time direction is not only important for
low-temperature calculations. The stability requirement for a standard Trotterization of the kinetic term can also be saturated at fixed temperature by increasing the spatial resolution and hence the sampling of high-momentum modes. A similar feature emerges when simulating Bose fluids in an optical trap, which is necessary for precise correspondence with ultracold atom experiments. The guaranteed linear stability of the present algorithm is therefore of significant practical value in virtually every CSCL simulation.

\section{Conclusion}
We conclude by emphasizing that the approach outlined here is complementary to many other techniques for taming numerical path integrals. Our approach can be used in higher-order decompositions \cite{zillichExtrapolatedHighorderPropagators2010, chinGradientSymplecticAlgorithms2002} of $e^{-\Delta\hat H}$ wherever one seeks to apply a quadratic operator to a coherent state. Furthermore, our treatment of quadratic propagators may enable a more robust incorporation of quadratic constraints, such as fixed particle number in the canonical ensemble \cite{mcgarrigleProjectedComplexLangevin2024}, by performing the projections within the path integral. 

Boson models with strong interactions still present a formidable challenge for CSCL approaches with our improved propagator method. Strong pairwise interactions produce frequent nonlinear numerical instabilities that result in method failure. However, a potential path forward would first decouple the quartic interaction via auxiliary fields from a Hubbard-Stratonovich transformation, and then proceed with Strang splitting and our quadratic propagator procedure. The resulting hybrid coherent state - auxiliary field theory would be second-order accurate and may exhibit superior numerical stability when sampling strongly interacting boson models. Finally we note that this approach is not limited to decompositions of imaginary-time propagators, and indeed can be readily applied to real-time contours in order to simulate  quantum many-body dynamics (via Keldysh contours) \cite{kamenevFieldTheoryNonEquilibrium2011, fredricksonFieldTheoreticSimulationsSoft2023} or classical many-body dynamics in the Doi-Peliti coherent states representation \cite{doiSecondQuantizationRepresentation1976, pelitiPathIntegralApproach1985,  fixmanPolymerDynamicsBoson1966}.

\section*{Acknowledgements}
This work was enabled by field-theoretic simulation tools developed under support from the National Science Foundation (CMMT Program, DMR-2104255). Use was made of computational facilities purchased with funds from the NSF (CNS-1725797) and administered by the Center for Scientific Computing (CSC). This work made use of the BioPACIFIC Materials Innovation Platform computing resources of the National Science Foundation Award No. DMR-1933487. The CSC is supported by the California NanoSystems Institute and the Materials Research Science and Engineering Center (MRSEC; NSF DMR 2308708) at UC Santa Barbara. E.C.M acknowledges support from a Mitsubishi Chemical Fellowship. 
TGK acknowledges support from the National Science Foundation under grant PHY-2309135 to the Kavli Institute for Theoretical Physics (KITP), and from the Gordon and Betty Moore Foundation through Grant GBMF8690 to the University of California, Santa Barbara.

\appendix

\section{Quadratic propagation of coherent state wavefunction}
Here we show that $e^{-\Delta\hat H_0/2}|\phi\rangle$, where $\hat H_0$ is a quadratic (bosonic) Hamiltonian and $|\phi\rangle$ is a coherent state, may be expressed as another coherent state, $|\phi^\prime\rangle$, with transformed complex field variables. We start by defining 
the single-particle eigenbasis of $\hat H_0$: $\hat H_0|\lambda\rangle=\epsilon_\lambda|\lambda\rangle$ with $|\lambda\rangle\equiv \hat a^\dagger_\lambda|0\rangle$. This defines a unitary matrix $U^\dagger_{\alpha,\lambda}=\langle\alpha|\lambda\rangle$ that diagonalizes the single-particle Hamiltonian. We then rewrite $|\phi\rangle$ in the eigenbasis of $\hat H_0$: $\exp(\sum_\lambda\phi(\lambda)\hat a^\dagger_\lambda)|0\rangle$. Note that this change of basis does not alter the form of the coherent-state wavefunction:
\begin{eqnarray}
        \sum_\alpha\phi(\alpha)\hat a^\dagger_\alpha&=&\sum_{\alpha,\lambda_1,\lambda_2}\phi(\lambda_1)U_{\lambda_1,\alpha}U^\dagger_{\alpha,\lambda_2}\hat a^\dagger_{\lambda_2}\\
        &=&\sum_{\lambda_1,\lambda_2}\phi(\lambda_1)\delta_{\lambda_1,\lambda_2}\hat a^\dagger_{\lambda_2}\\
        &=&\sum_\lambda\phi(\lambda)\hat a^\dagger_\lambda
\end{eqnarray}
The transformed state is defined as
\begin{equation}
    |\phi^\prime\rangle=e^{-\Delta\hat H_0/2}\exp\left(\sum_\lambda\phi(\lambda)\hat a^\dagger_\lambda\right)|0\rangle
\end{equation}
We now use the condition of a trivial vacuum state, $e^{-\Delta\hat H_0/2}|0\rangle=|0\rangle$, 
to insert an inverse kinetic propagator in front of the vacuum state:
\begin{equation}
    |\phi^\prime\rangle=e^{-\Delta\hat H_0/2}\exp\left(\sum_\lambda\phi(\lambda)\hat a^\dagger_\lambda\right)e^{\Delta\hat H_0/2}|0\rangle.
\end{equation}
We may pull the factors of $e^{\pm\Delta\hat H_0/2}$ into the exponential:
\begin{equation}
    |\phi^\prime\rangle=\exp\left(e^{-\Delta\hat H_0/2}\sum_\lambda\phi(\lambda)\hat a^\dagger_\lambda e^{\Delta\hat H_0/2}\right)|0\rangle
\end{equation}
Finally, using the second-quantized form $\hat H_0=\sum_\lambda\epsilon_\lambda n_\lambda$, the propagators may be commuted through the sum to find:
\begin{equation}
    |\phi^\prime\rangle=\exp\left(\sum_\lambda e^{-\Delta\epsilon_\lambda/2}\phi(\lambda)\hat a^\dagger_\lambda \right)|0\rangle
\end{equation}
Thus we have shown that $|\phi^\prime$ is a coherent state wavefunction that is related to $|\phi\rangle$ by a transformation of the complex field variables in the eigenbasis of $\hat H_0$: $\phi^\prime(\lambda)=e^{-\Delta\epsilon_\lambda/2}\phi(\lambda)$. Following these same steps, it is straightforward to derive that $\langle\phi|e^{-\Delta\hat H_0/2}=\langle\phi^\prime|$ is a coherent state defined by the transformation $(\phi^*(\lambda))^\prime=e^{-\Delta\epsilon_\lambda/2}\phi^*(\lambda)$. 

{The case of non-interacting bosons provides a useful benchmark that highlights the efficacy of this approach. In the non-interacting limit ($g = 0$), no Strang splitting error is incurred nor is Trotterization required, as the kinetic propagator $e^{-\beta \hat{H}_{0}}$ can be applied exactly to all orders in $\beta$ using the kinetic-propagated basis $\ket{\phi'}$. As a result, the ideal gas partition function is obtained after performing Gaussian integrals over the coherent state amplitudes \emph{without invoking the usual $N_{\tau} \to \infty$ limit} \cite{negeleQuantumManyparticleSystems1988}. As a result, simulations of non-interacting bosons show no $N_{\tau}$ dependence in their observables, underscoring the exactness of our approach. }

\begin{figure*}
    \centering
    \includegraphics[scale = 0.72]{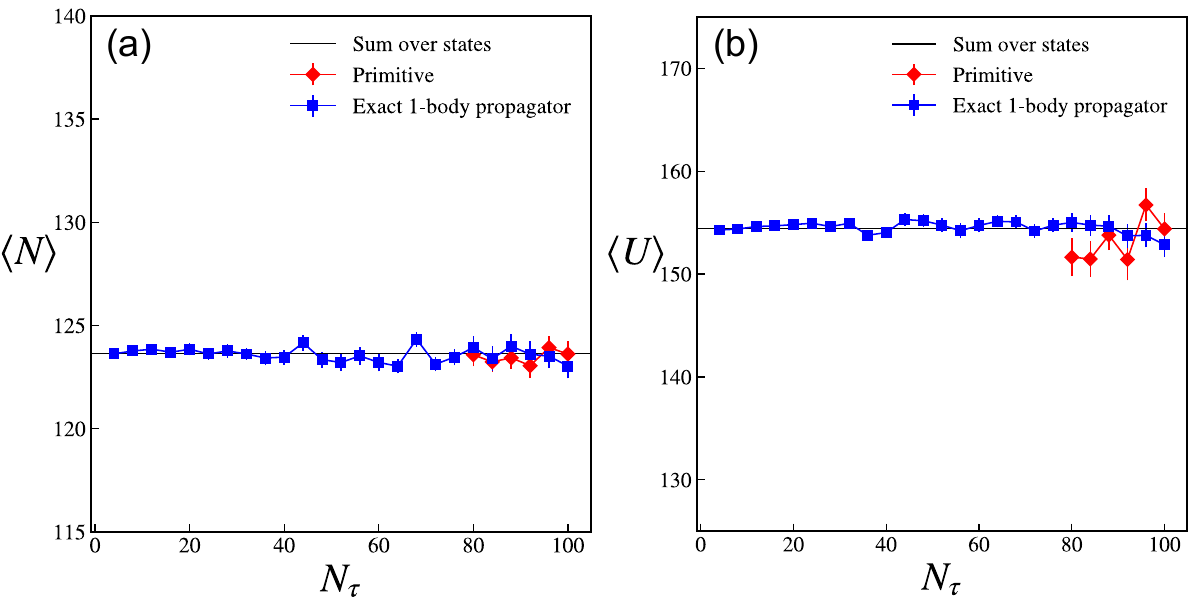}
    \caption{Imaginary time dependence of ideal Bose gas simulations. Ensemble-averaged a) particle number and b) internal energy as a function of imaginary time discretization $N_{\tau}$ for the primitive method (red diamonds) and the new exact propagator method (blue squares) introduced in this manuscript. The black solid line denotes the exact thermodynamic reference for the ideal Bose gas, computed using a sum over wavevector basis states with sufficient momentum cutoff. Data from the primitive data below $N_{\tau} = 80$ are not shown due to numerical instability, which prevents meaningful data collection. Simulations consisted of a two-dimensional ensemble of non-interacting bosons with $\bar{g}_{\rm 2D} = 0$, $\ell = \sqrt{\hbar^2 /2m |\mu|} = 4.29$, $\mu = -0.33$ K, $\beta = 0.50$ K$^{-1}$, $L_{x} / \ell = L_{y} /\ell = 11.67$, and $N_{x} = N_{y} = 80$ grid points in each direction. The exponential-time-differencing (ETD) numerical algorithm was used with a Langevin step discretization $\Delta t = 0.0033$. Error bars depict standard errors of the mean, determined during sample averaging.}
    \label{fig: ideal_gas_SI}
\end{figure*}

{Shown in Figure\ (\ref{fig: ideal_gas_SI}), the particle number and internal energy observables are independent of the imaginary time discretization and enjoy absolute numerical stability. We readily compare results for non-interacting bosons with known thermodynamic references, which are computed by an explicit summation over wavevector states: }
\begin{eqnarray}
    \langle N \rangle &=& \sum_{\mathbf{k}}\frac{1}{e^{\beta (\epsilon_{\mathbf{k}} - \mu)} -1}\cr\cr
    \langle U \rangle &=& \sum_{\mathbf{k}}\frac{\epsilon_{\mathbf{k}}}{e^{\beta (\epsilon_{\mathbf{k}} - \mu)} -1} 
    \label{eq: sum_over_states_idealgas}
\end{eqnarray}
{\noindent{where} $\epsilon_{\mathbf{k}} = \hbar^2 \mathbf{k}^2/2m$ is the continuum kinetic dispersion and a suitably large momentum cutoff is chosen such that the sum converges. The thermodynamic results of equations\ (\ref{eq: sum_over_states_idealgas}) are derived from the partition function and are valid in any spatial dimensionality. The agreement of both methods with the analytical reference serves to simultaneously validate our exact propagator method while highlighting its favorable numerical stability and accuracy properties. }

\section{Extension to Pseudospin-1/2 Coherent States with Rashba Spin-orbit Coupling}
Here we detail the applicability of our approach to multicomponent systems with a non-trivial one-body spectra. Using two-photon Raman transitions coupled with a biasing magnetic field, it is possible to create a system Pseudospin-1/2 bosons from a $S=1$ boson system \cite{linSpinOrbitCoupled2011}. Furthermore, with additional Raman lasers \cite{campbellRashbaRealizationRaman2016, valdes-curielTopologicalFeaturesLattice2021}, it is possible to imbue the system with a two-dimensional isotropic ``Rashba'' spin-orbit coupling. The Rashba coupling constitutes a fascinating case of great fundamental interest due to the massive single particle degeneracy of the Rashba Hamiltonian. Here, a degenerate ring of states appears with radius $|\mathbf{k}| = \kappa$, with $\kappa$ taken as the spin-orbit coupling strength. The SOC Hamiltonian can be expressed in a compact vectorized form: 
\begin{equation}
    \hat{H}_{\text{SOC}} = \int d^2 r\ \hat{\Psi}^{\dagger} \left [ \frac{1}{2m} (\hat{\bm{p}} \underline{\underline{I}} - \bm{\mathcal{A}})^2 \right ] \hat{\Psi}
\end{equation}
\noindent{where} $\hat{\Psi} = ( \hat{\psi}_{\uparrow} (\mathbf{r}) , \hat{\psi}_{\downarrow} (\mathbf{r}) )^{T}$ is a two-component vector of second-quantized field operators satisfying Bose commutation relations \cite{negeleQuantumManyparticleSystems1988}.

After rescaling by the healing length as in reference \cite{mcgarrigleEmergenceSpinMicroemulsion2023} and incorporating the necessary chemical potential shift $\mu'$ to ensure a positive definite spectrum, the resulting rank-2 Rashba Hamiltonian matrix can be written as $\hat{H}_{\text{SOC}} / \mu_{\text{eff}} = \sum_{\mathbf{k}} \hat{\Psi}^{\dagger}_{\mathbf{k}} K_{\mathbf{k}} \hat{\Psi}_{\mathbf{k}}$. The one-body matrix $K_{\mathbf{k}}$ has diagonal elements $K_{\alpha \alpha} = \mathbf{k}^2 + \mu'$ for both $\alpha = \uparrow, \downarrow$ and off-diagonal elements $K_{\uparrow \downarrow} = -2\kappa |\mathbf{k}| e^{-i\theta_{\mathbf{k}}}$ and $K_{\downarrow \uparrow} = K^*_{\uparrow \downarrow}$, with $\theta_{\mathbf{k}} = \arctan(k_{y}/k_{x})$. This can be diagonalized by a unitary transformation $U_{\mathbf{k}}$ to yield $D_{\mathbf{k}} = U_{\mathbf{k}} K_{\mathbf{k}} U^{\dagger}_{\mathbf{k}}$, where $D_{\mathbf{k}}$ is a diagonal matrix with entries $D_{00} = E_{+}(\mathbf{k})$ and $D_{11} = E_{-}(\mathbf{k})$. The unitary transformation involves eigenvectors $U_{\mathbf{k}} = [v_{+, \mathbf{k}} , v_{-,\mathbf{k}}]$, which are defined $v_{+, \mathbf{k}} = 2^{-1/2} (1, -\exp(\theta_{\mathbf{k}}))^{T}$ and $v_{-, \mathbf{k}} = 2^{-1/2}(1, \exp(\theta_{\mathbf{k}}))^{T}$.  The eigenvalue spectrum is $E_{\pm}(\mathbf{k}) = \mathbf{k}^2 \pm 2\kappa |\mathbf{k}| + \mu'$. Thus, the Hamiltonian is easily re-expressed a $\hat{H}_{\text{SOC}} / \mu_{\text{eff}} = \sum_{\mathbf{k}} \tilde{\Psi}^{\dagger}_{\mathbf{k}} D_{\mathbf{k}} \tilde{\Psi}_{\mathbf{k}}$, where we have switched to a diagonal ``dressed state'' field operator basis: $\tilde{\Psi}_{\mathbf{k}} = U_{\mathbf{k}} \hat{\Psi}_{\mathbf{k}}$ and $\tilde{\Psi}^{\dagger}_{\mathbf{k}} = \hat{\Psi}_{\mathbf{k}}^{\dagger} U^{\dagger}_{\mathbf{k}}$. Instead of pseudospin states, the field operator components represent helicity states $\ket{\pm;\ \mathbf{k}}$ that correspond with the bands $E_{\pm}(\mathbf{k})$.  

When building the coherent state path integral for Rashba bosons, the unitary transformation $U_{\mathbf{k}}$ yields dressed coherent state vectors $\bm{\tilde{\phi}}_{\mathbf{k}} = U_{\mathbf{k}} \bm{\phi_{\mathbf{k}}}$ and $\bm{\tilde{\phi}^{\dagger}}_{\mathbf{k}} = \bm{\phi_{\mathbf{k}} ^{\dagger}} U^{\dagger}_{\mathbf{k}}$ with no change in the functional integration measure, identity resolution, or coherent state amplitudes: $\ket{\phi_{j}} = \exp(\sum_{\mathbf{k}}  \bm{\hat{b}}^{\dagger}_{\mathbf{k}} \bm{\phi}_{\mathbf{k}}) \ket{0} = \exp(\sum_{\mathbf{k}} \bm{\hat{b}} ^{\dagger}_{\mathbf{k}}  U_{\mathbf{k}}  U^{\dagger}_{\mathbf{k}} \bm{\phi}_{\mathbf{k}})\ket{0} = \exp(\sum_{\mathbf{k}} \bm{\tilde{b}} ^{\dagger}_{\mathbf{k}} \bm{\tilde{\phi}}_{\mathbf{k}})\ket{0}$. We apply the Strang splitting for the propagator in each matrix element $\langle\phi_j|e^{-\Delta \hat{H}}|\phi_{j-1}\rangle = \langle\phi_j|e^{-\Delta \hat{H}_{\text{SOC}}/2} e^{-\Delta \hat{H}_{\text{int}}}e^{-\Delta \hat{H}_{\text{SOC}}/2}|\phi_{j-1}\rangle + \mathcal{O}(\Delta^3_{\tau})$ and then evaluate the SOC propagator on the the coherent state exactly: 
\begin{eqnarray*}
    e^{-\Delta \hat{H}_{\text{SOC}}/2}\ket{\phi_{j}} &=& e^{-\Delta/2 \sum_{\mathbf{k}} \tilde{\Psi}^{\dagger}_{\mathbf{k}} D_{\mathbf{k}} \tilde{\Psi}_{\mathbf{k}} } \ket{\phi_{j}}\\
    &=& e^{-\Delta/2 \sum_{\mathbf{k}} \tilde{\Psi}^{\dagger}_{\mathbf{k}} D_{\mathbf{k}} \tilde{\Psi}_{\mathbf{k}} } e^{\sum_{\mathbf{k}} \bm{\tilde{b}}^{\dagger}_{\mathbf{k}} \bm{\tilde{\phi}}_{\mathbf{k}}}\ket{0} \\
    &=& \exp(\sum_{\mathbf{k}}\bm{\tilde{b}}^{\dagger}_{\mathbf{k}} e^{-\Delta D_{\mathbf{k}}/2} \bm{\tilde{\phi}}_{\mathbf{k}}\ket{0} \\
    &\equiv& \ket{\phi'_{j}}
\end{eqnarray*}
\noindent{where} we have safely leveraged the previous insights since the dressed bosons obey the usual commutation relations $[\hat{b}_{\alpha', \mathbf{k}}, \hat{b}^{\dagger}_{\gamma', \mathbf{k}'}]= \delta_{\mathbf{k}, \mathbf{k}'} \delta_{\alpha', \gamma'}$, with $\alpha'$ denoting a dressed state index. 

Therefore, we can proceed, but care must be taken with the interaction Hamiltonian, which is composed of density-density interactions written in the pseudospin basis. 
\begin{eqnarray}
    \langle\phi_j|e^{-\Delta \hat{H}}|\phi_{j-1}\rangle = \langle\phi'_j|e^{-\Delta \hat{H}_{\text{int}}}|\phi'_{j-1}\rangle 
\end{eqnarray}
In the usual first-order Taylor expansion, we encounter terms where operators in the pseudospin basis must be evaluated with dressed coherent states, such as $\langle\phi'_j| \hat{b}^{\dagger}_{\gamma} \hat{b}^{\dagger}_{\alpha} \hat{b}_{\gamma} \hat{b}_{\alpha}|\phi'_{j-1}\rangle$. We evaluate them as such: 
\begin{eqnarray}
    \hat{b}_{\alpha, \mathbf{k}} \ket{\phi'_{j}} &=& \sum_{\gamma'}(U_{\mathbf{k}})_{\alpha \gamma'} \tilde{b}_{\gamma', \mathbf{k}}\ket{\phi'_{j}} \\
    &=& \sum_{\gamma'}(U_{\mathbf{k}})_{\alpha \gamma'} \tilde{\phi}'_{\gamma', j, \mathbf{k}}\ket{\phi'_{j}} \\
    &=& \phi'_{\alpha, j, \mathbf{k}}\ket{\phi'_{j}}
\end{eqnarray}
\noindent{where} we have defined a coherent state amplitude in the pseudospin basis that's been propagated forward with the kinetic energy operator. 

Now, we show that the field operator in the pseudospin basis acting on $\ket{\phi'_{j}}$ yields the following:
\begin{eqnarray}
    \hat{\psi}_{\alpha} (\mathbf{r}) \ket{\phi'_{j}} &=& \sum_{\mathbf{k}} \psi_{\alpha}(\mathbf{r}) b_{\alpha, \mathbf{k}} \ket{\phi'_{j}}\\
    &=& \sum_{\mathbf{k}} \psi_{\alpha}(\mathbf{r})\phi'_{\alpha, j, \mathbf{k}}\ket{\phi'_{j}} \\
    &=& \phi'_{\alpha, j} (\mathbf{r})\ket{\phi'_{j}},
\end{eqnarray}
\noindent{where} in practice we would compute this propagated pseudospin coherent state via the following procedure:
\begin{eqnarray}
    \bm{\phi}'_{j}(\mathbf{r}) &=& \mathcal{F}^{-1}_{\mathbf{k} \to \mathbf{r}}\left [ U_{\mathbf{k}} (e^{-\Delta D_{\mathbf{k}}/2}\tilde{\bm{\phi}}) \right ] \\
    &=& \mathcal{F}^{-1}_{\mathbf{k} \to \mathbf{r}}\left [ U_{\mathbf{k}} (e^{-\Delta D_{\mathbf{k}}/2} U^{\dagger}_{\mathbf{k}} \mathcal{F}_{\mathbf{r} \to \mathbf{k}}[\bm{\phi}_{j} (\mathbf{r})]) \right ]
\end{eqnarray}
\noindent{with} $\mathcal{F}$ $(\mathcal{F}^{-1})$ denoting the Fourier transform in the forward (backward) direction, with the following convention for a function $f(\mathbf{r})$: $\mathcal{F}_{\mathbf{r} \to \mathbf{k}}[f] = \frac{1}{V} \int d \mathbf{r}\ \exp(-i \mathbf{k} \cdot \mathbf{r}) f(\mathbf{r})$. 

Collecting all factors, we arrive at a partition function $\mathcal{Z} = \int \mathcal{D} (\bm{\tilde{\phi}}, \bm{\tilde{\phi}^{\dagger}})\ e^{-S[\bm{\tilde{\phi}}, \bm{\tilde{\phi}^{\dagger}}]}$ with the discrete imaginary time action for the Rashba boson system $S[\bm{\tilde{\phi}}, \bm{\tilde{\phi}^{\dagger}}] = S_{0}[\bm{\phi}, \bm{\phi^{\dagger}}] + S_{\rm int}[(\bm{\phi}, \bm{\phi^{\dagger}})']$:
\begin{equation}
    S_{0}[\bm{\tilde{\phi}}, \bm{\tilde{\phi}^{\dagger}}] = \sum_{\mathbf{k}} \sum_{j=0}^{N_{\tau}-1} \bm{\tilde{\phi}}^{\dagger}_{j, \mathbf{k}}[\bm{\tilde{\phi}}_{j, \mathbf{k}} - e^{-\Delta D_{\mathbf{k}}}\bm{\tilde{\phi}}_{j-1, \mathbf{k}}]
\end{equation}
\begin{multline}
    S_{\text{int}}[(\bm{\phi}, \bm{\phi^{\dagger}})'] = \\
    \frac{\Delta}{2} \sum_{j=0}^{N_{\tau}-1}\int d\mathbf{r}\sum_{\alpha \gamma} g_{\alpha \gamma}\phi^*_{\alpha, j} (\mathbf{r})' \phi^*_{\gamma, j} (\mathbf{r})' \phi'_{\alpha, j-1} (\mathbf{r})' \phi'_{\gamma, j-1} (\mathbf{r})
\end{multline}
\noindent{which} is studied in the main text. 

As discussed in the main text, it is advantageous to conduct CL in the diagonalized basis. Since the unitary transformation does not modify the functional integral measure, we can prescribe the same off-diagonal descent Langevin scheme for coherent states in the dressed basis: 
\begin{eqnarray}
    \frac{\partial \bm{\tilde{\phi}}_{j, \mathbf{\mathbf{k}}}}{\partial t} &= -\frac{\delta S[\bm{\tilde{\phi}}, \bm{\tilde{\phi}^*}]}{\delta \bm{\tilde{\phi}^*}_{j, \mathbf{k}}} + \bm{\eta}_{j, \mathbf{k}} \\
    \frac{\partial \bm{\tilde{\phi}^*}_{j, \mathbf{\mathbf{k}}}}{\partial t} &= -\frac{\delta S[\bm{\tilde{\phi}}, \bm{\tilde{\phi}^*}]}{\delta \bm{\tilde{\phi}}_{j, \mathbf{k}}} + \bm{\eta^*}_{j, \mathbf{k}}
\end{eqnarray} 
\noindent{where} $\bm{\eta}_{j, \mathbf{k}}$ and $\bm{\eta}^*_{j, \mathbf{k}}$ are diagonal, complex-conjugate noise vectors which are constructed in real space with two independent real noise source fields: $\eta_{\alpha, j} (\mathbf{r}, t) = \eta^{(R)}_{\alpha,j}(\mathbf{r},t ) + i\eta^{(I)}_{\alpha,j}(\mathbf{r}, t)$ and $\eta^*_{\alpha, j} (\mathbf{r},t ) = \eta^{(R)}_{\alpha,j}(\mathbf{r},t ) - i\eta^{(I)}_{\alpha,j}(\mathbf{r}, t)$, where the real and imaginary parts are generated independently with the following statistics $\langle \eta^{A}_{\alpha, j} (\mathbf{r}, t) \eta^{B}_{\gamma, j'} (\mathbf{r'}, t')\rangle = \delta_{j, j'} \delta_{A, B} \delta_{\alpha, \gamma} \delta(\mathbf{r} - \mathbf{r'})\delta(t - t')$. 

Importantly, the interaction part of the action is a function of propagated fields in the pseudospin basis, so we provide clarification on how to obtain the correct expressions for the thermodynamic forces required for Langevin sampling. For the $\bm{\phi}$ CL equation, the functional derivatives for the $S_{\rm int}$ portion of the action are determined by carefully including the appropriate Jacobian factors and transformations: 
\begin{eqnarray*}
\frac{\delta S_{\text{int}}[(\bm{\phi}, \bm{\phi^*})']}{\delta \bm{\tilde{\phi}^*}_{j, \mathbf{k}}} &=& e^{-\Delta D_{\mathbf{k}}/2} \left (\frac{\delta S_{\text{int}}[(\bm{\phi}, \bm{\phi^*})']}{\delta (\bm{\tilde{\phi}^*}_{j, \mathbf{k}})'} \right )\\
&=&  e^{-\Delta D_{\mathbf{k}}/2} \left ( \frac{\delta (\bm{\phi^*}_{\mathbf{k}})'}{\delta (\bm{\tilde{\phi}^*}_{\mathbf{k}})'} \right)\left (\frac{\delta S_{\text{int}}[(\bm{\phi}, \bm{\phi^*})']}{\delta (\bm{\phi^*}_{j, \mathbf{k}})'} \right ) \\ &=& e^{-\Delta D_{\mathbf{k}}/2} U_{\mathbf{k}} \left (\frac{\delta S_{\text{int}}[(\bm{\phi}, \bm{\phi^*})']}{\delta (\bm{\phi^*}_{j, \mathbf{k}})'} \right ) \\
&=& e^{-\Delta D_{\mathbf{k}}/2} U_{\mathbf{k}} \mathcal{F}_{\mathbf{r \to \mathbf{k}}} \left [ \frac{\delta S_{\text{int}}[(\bm{\phi}, \bm{\phi^*})']}{\delta \bm{\phi^*}_{j}(\mathbf{r})'}    \right ]
\end{eqnarray*}
\noindent{where} we have used the Jacobian $U_{\mathbf{k}}$ for the dressed state transformation. For the force on $\phi^*$, we find the following: 
\begin{eqnarray*}
\frac{\delta S_{\text{int}}[(\bm{\phi}, \bm{\phi^*})']}{\delta \bm{\tilde{\phi}}_{j, \mathbf{k}}} &=&  e^{-\Delta D_{\mathbf{k}}/2} \left (\frac{\delta S_{\text{int}}[(\bm{\phi}, \bm{\phi^*})']}{\delta (\bm{\tilde{\phi}}_{j, \mathbf{k}})'} \right ) \\ &=&  e^{-\Delta D_{\mathbf{k}}/2} \left (\frac{\delta S_{\text{int}}[(\bm{\phi}, \bm{\phi^*})']}{\delta (\bm{\phi}_{j, \mathbf{k}})'} \right ) \left ( \frac{\delta (\bm{\phi}_{\mathbf{k}})'}{\delta (\bm{\tilde{\phi}}_{\mathbf{k}})'} \right) \\
 &=& e^{-\Delta D_{\mathbf{k}}/2} \left (\frac{\delta S_{\text{int}}[(\bm{\phi}, \bm{\phi^*})']}{\delta (\bm{\phi}_{j, \mathbf{k}})'} \right ) U^{\dagger}_{\mathbf{k}}  \\
&=& \mathcal{F}_{\mathbf{r \to \mathbf{k}}} \left [ \frac{\delta S_{\text{int}}[(\bm{\phi}, \bm{\phi^*})']}{\delta \bm{\phi}_{j}(\mathbf{r})'} \right ] U^{\dagger}_{\mathbf{k}} e^{-\Delta D_{\mathbf{k}}/2}.
\end{eqnarray*}
\noindent{From} here, the usual expressions for the functional derivative of the interacting portion of the action from reference\ (\cite{delaneyNumericalSimulationFiniteTemperature2020, fredricksonFieldTheoreticSimulationsSoft2023}) may be used.

\section{Estimation of Observables}
Under the exact quadratic propagator treatment, computing observables during Langevin sampling requires a modified procedure. A central finding of this work is that observables should be computed using the propagated coherent state fields $(\phi, \phi^*)'$ with the functionals derived previously in the primitive first-order approach \cite{fredricksonFieldTheoreticSimulationsSoft2023}. In other words, the functionals from the primitive approach are reused with the substitution $(\phi, \phi^*) \to (\phi, \phi^*)'$, enabling straightforward implementation in existing software. Below we discuss the supporting arguments for this conclusion. 

For a local observable $\hat{O}(\mathbf{ r})$, the procedure for developing an estimator functional of the coherent state fields $\tilde{O}[\phi, \phi^*; \mathbf{r}]$ involves augmenting the Hamiltonian with an infinitesimal source field $J(\mathbf{r})$ conjugate to the observable, so that $\hat{H} \to \hat{H} + \hat{H}_{J}$ with $\hat{H}_{J} = -\beta^{-1} \int d\mathbf{r}\ J(\mathbf{r}) \hat{O}(\mathbf{r})$ \cite{fredricksonFieldTheoreticSimulationsSoft2023}. When retracing the steps to derive the field theory, one could choose to incorporate this source Hamiltonian with the interaction portion, leading to the matrix elements $\langle\phi'_{j}|e^{-\Delta \hat{H}_{\text{int}}} e^{-\Delta \hat{H}_{J}} | \phi'_{j-1} \rangle $ to compute. Evaluating via a first-order Taylor expansion, we find that the operator functional would be evaluated using coherent state fields in the propagated basis, e.g. $\tilde{O}[(\phi, \phi^*)] = O[(\phi, \phi^*)']$. For example, a one-body observable $\hat{O}$ is written using second quantized operators $\hat{O} = \int d\mathbf{r}\ \psi^{\dagger}(\mathbf{r}) O(\mathbf{r}) \hat{\psi}(\mathbf{r})$ \cite{fetterQuantumTheoryManyParticle2012} and can be directly evaluated via this first-order accurate procedure. For instance, the density operator $\hat{\rho}(\mathbf{r}) = \psi^{\dagger}(\mathbf{r}) \psi^{\vphantom{\dagger}}(\mathbf{r})$ leads to the density operator expression: 
\begin{eqnarray}
 \frac{1}{\mathcal{Z}} \frac{\delta \mathcal{Z}[J]}{\delta J(\mathbf{r})}\biggr|_{J=0} &=& \langle \tilde{\rho}[\phi, \phi^*; \mathbf{r}] \rangle \\ \nonumber\\
 \tilde{\rho}[\phi, \phi^*; \mathbf{r}] &=& \frac{1}{N_{\tau}}\sum_{j=0}^{N_{\tau}-1}\phi^*_{j} (\mathbf{r})'\phi'_{j-1}(\mathbf{r})
\end{eqnarray}
\noindent{Spatial integration} of this density functional recovers the expression for the total particle number from the main text, i.e. $\tilde{N}[\phi, \phi^*] = \int d\mathbf{r}\ \tilde{\rho}[\phi, \phi^*; \mathbf{r}]$, which is additionally verifiable by taking $-\frac{\partial S}{\partial \mu}$. This result implies that for a general operator that can be derived in this form -- 
using an infinitesimal source field $J(\mathbf{r})$ -- 
the operator can be computed using the same expressions as before using the propagated coherent state fields instead. We emphasize that we additionally observe excellent accuracy at modest imaginary time discretizations for more complex observables, such as the superfluid density, pressure, and density structure factor. 


\section{Comparison to mean-field theory}
\begin{figure*}
    \centering
    \includegraphics[scale=0.56]{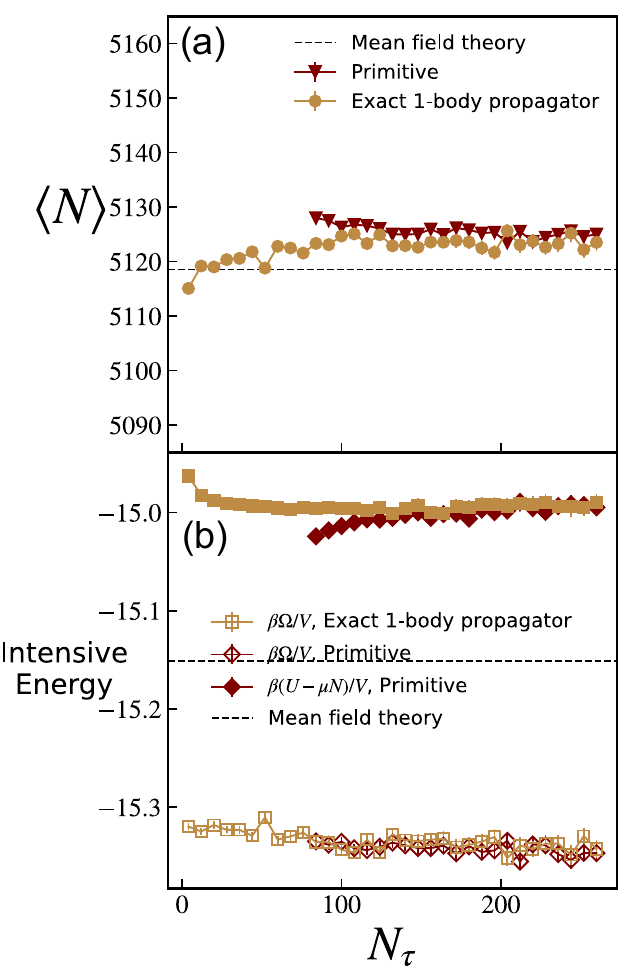}
    \includegraphics[scale=0.52]{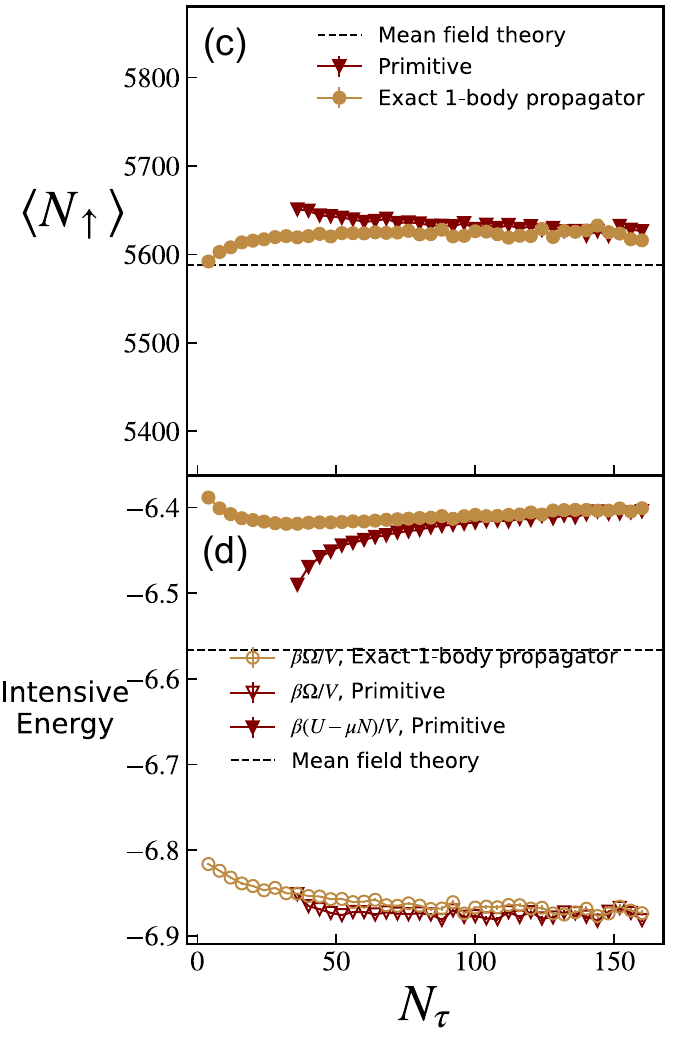}
    \caption{Comparison of the results from the main text with mean-field theory references for both the a), b) single-component homogeneous Bose gas and the c), d) Rashba spin-orbit coupled Bose gas in the stripe phase. The mean-field reference for the corresponding observables are plotted in the black dashed line in all cases.}
    \label{fig: SI_MFT}
\end{figure*}

{Although both examples in our study are qualitatively in the mean-field regime with a high condensate fraction, both stochastic methods considered here are beyond mean-field in nature, capturing the impact of fluctuations on observables. Both the ``primitive'' propagator approach and our exact ``quadratic propagator'' method provide exact estimates of observables' expectation values and include both quantum and thermal fluctuations \cite{fredricksonFieldTheoreticSimulationsSoft2023}. A Gross-Pitaevski \cite{pitaevskiiBoseEinsteinCondensationSuperfluidity2016} or mean-field approach assumes one dominant configuration in the partition function, such that the BEC is governed by a semi-classical energy functional of the BEC macroscopic wavefunction. In Figure\ (\ref{fig: SI_MFT}), we plot the estimate of each observable via mean-field theory, which assumes a fully condensed system at zero temperature. Thermodynamically, mean-field theory assumes the following relationship between the internal energy and the grand free energy: $\langle U \rangle  - \mu \langle N \rangle = \langle \Omega \rangle $. Both Figure\ (\ref{fig: SI_MFT}b) and (\ref{fig: SI_MFT}d) show that this equality is violated, as expected at non-zero temperature. Beyond the scope of this study, however, would be a comparison of the mean-field results with a $T=0$ ground state method that includes quantum fluctuations. }

\bibliography{CSCL_Propagator_Methods} 

@article{attanasioThermodynamicsSpinorbitcoupledBosons2020,
  title = {Thermodynamics of Spin-Orbit-Coupled Bosons in Two Dimensions from the Complex {{Langevin}} Method},
  author = {Attanasio, Felipe and Drut, Joaqu{\'i}n E.},
  year = {2020},
  month = mar,
  journal = {Physical Review A},
  volume = {101},
  number = {3},
  pages = {033617},
  publisher = {American Physical Society},
  doi = {10.1103/PhysRevA.101.033617},
  urldate = {2022-07-22}
}

@article{blandTwoDimensionalSupersolidFormation2022,
  title = {Two-{{Dimensional Supersolid Formation}} in {{Dipolar Condensates}}},
  author = {Bland, T. and Poli, E. and Politi, C. and Klaus, L. and Norcia, M. A. and Ferlaino, F. and Santos, L. and Bisset, R. N.},
  year = {2022},
  month = may,
  journal = {Physical Review Letters},
  volume = {128},
  number = {19},
  pages = {195302},
  issn = {0031-9007, 1079-7114},
  doi = {10.1103/PhysRevLett.128.195302},
  urldate = {2022-05-16},
  langid = {english}
}

@article{blasiakCombinatoricsBosonNormal2007,
  title = {Combinatorics and {{Boson}} Normal Ordering: {{A}} Gentle Introduction},
  shorttitle = {Combinatorics and {{Boson}} Normal Ordering},
  author = {Blasiak, P. and Horzela, A. and Penson, K. A. and Solomon, A. I. and Duchamp, G. H. E.},
  year = {2007},
  month = jul,
  journal = {American Journal of Physics},
  volume = {75},
  number = {7},
  pages = {639--646},
  issn = {0002-9505},
  doi = {10.1119/1.2723799},
  urldate = {2025-06-24}
}

@article{campbellRashbaRealizationRaman2016,
  title = {Rashba Realization: {{Raman}} with {{RF}}},
  shorttitle = {Rashba Realization},
  author = {Campbell, D. L. and Spielman, I. B.},
  year = {2016},
  month = apr,
  journal = {New Journal of Physics},
  volume = {18},
  number = {3},
  pages = {033035},
  publisher = {IOP Publishing},
  issn = {1367-2630},
  doi = {10.1088/1367-2630/18/3/033035},
  urldate = {2023-03-03},
  langid = {english}
}

@article{capogrosso-sansoneMonteCarloStudy2008,
  title = {Monte {{Carlo}} Study of the Two-Dimensional {{Bose-Hubbard}} Model},
  author = {{Capogrosso-Sansone}, Barbara and S{\"o}yler, {\c S}ebnem G{\"u}ne{\c s} and Prokof'ev, Nikolay and Svistunov, Boris},
  year = {2008},
  month = jan,
  journal = {Physical Review A},
  volume = {77},
  number = {1},
  pages = {015602},
  issn = {1050-2947, 1094-1622},
  doi = {10.1103/PhysRevA.77.015602},
  urldate = {2022-06-12},
  langid = {english}
}

@article{ceperleyPathintegralComputationLowtemperature1986,
  title = {Path-Integral Computation of the Low-Temperature Properties of Liquid \${\textasciicircum}\{4\}{\textbackslash}mathrm\{\vphantom\}{{He}}\vphantom\{\}\$},
  author = {Ceperley, D. M. and Pollock, E. L.},
  year = {1986},
  month = jan,
  journal = {Physical Review Letters},
  volume = {56},
  number = {4},
  pages = {351--354},
  publisher = {American Physical Society},
  doi = {10.1103/PhysRevLett.56.351},
  urldate = {2024-06-01}
}

@article{ceperleyPathintegralSimulationSuperfluid1989,
  title = {Path-Integral Simulation of the Superfluid Transition in Two-Dimensional \${\textasciicircum}\{4\}{\textbackslash}mathrm\{\vphantom\}{{He}}\vphantom\{\}\$},
  author = {Ceperley, D. M. and Pollock, E. L.},
  year = {1989},
  month = feb,
  journal = {Physical Review B},
  volume = {39},
  number = {4},
  pages = {2084--2093},
  publisher = {American Physical Society},
  doi = {10.1103/PhysRevB.39.2084},
  urldate = {2024-06-01}
}

@article{ceperleyPathIntegralsTheory1995,
  title = {Path Integrals in the Theory of Condensed Helium},
  author = {Ceperley, D. M.},
  year = {1995},
  month = apr,
  journal = {Reviews of Modern Physics},
  volume = {67},
  number = {2},
  pages = {279--355},
  publisher = {American Physical Society},
  doi = {10.1103/RevModPhys.67.279},
  urldate = {2024-07-30}
}

@article{chinGradientSymplecticAlgorithms2002,
  title = {Gradient Symplectic Algorithms for Solving the {{Schr{\"o}dinger}} Equation with Time-Dependent Potentials},
  author = {Chin, Siu A. and Chen, C. R.},
  year = {2002},
  month = jul,
  journal = {The Journal of Chemical Physics},
  volume = {117},
  number = {4},
  pages = {1409--1415},
  issn = {0021-9606},
  doi = {10.1063/1.1485725},
  urldate = {2025-07-21}
}

@article{cooperTopologicalBandsUltracold2019,
  title = {Topological Bands for Ultracold Atoms},
  author = {Cooper, N. R. and Dalibard, J. and Spielman, I. B.},
  year = {2019},
  month = mar,
  journal = {Reviews of Modern Physics},
  volume = {91},
  number = {1},
  pages = {015005},
  publisher = {American Physical Society},
  doi = {10.1103/RevModPhys.91.015005},
  urldate = {2025-07-01}
}

@article{delaneyNumericalSimulationFiniteTemperature2020,
  title = {Numerical {{Simulation}} of {{Finite-Temperature Field Theory}} for {{Interacting Bosons}}},
  author = {Delaney, Kris T. and Orland, Henri and Fredrickson, Glenn H.},
  year = {2020},
  month = feb,
  journal = {Physical Review Letters},
  volume = {124},
  number = {7},
  pages = {070601},
  publisher = {American Physical Society},
  doi = {10.1103/PhysRevLett.124.070601},
  urldate = {2021-07-13}
}

@article{doiSecondQuantizationRepresentation1976,
  title = {Second Quantization Representation for Classical Many-Particle System},
  author = {Doi, M.},
  year = {1976},
  month = sep,
  journal = {Journal of Physics A: Mathematical and General},
  volume = {9},
  number = {9},
  pages = {1465},
  issn = {0305-4470},
  doi = {10.1088/0305-4470/9/9/008},
  urldate = {2025-08-01},
  langid = {english}
}

@book{fetterQuantumTheoryManyParticle2012,
  title = {Quantum {{Theory}} of {{Many-Particle Systems}}},
  author = {Fetter, Alexander L. and Walecka, John Dirk},
  year = {2012},
  month = mar,
  publisher = {Courier Corporation},
  googlebooks = {t5\_DAgAAQBAJ},
  isbn = {978-0-486-13475-8},
  langid = {english}
}

@article{feynmanSpaceTimeApproachNonRelativistic1948,
  title = {Space-{{Time Approach}} to {{Non-Relativistic Quantum Mechanics}}},
  author = {Feynman, R. P.},
  year = {1948},
  month = apr,
  journal = {Reviews of Modern Physics},
  volume = {20},
  number = {2},
  pages = {367--387},
  publisher = {American Physical Society},
  doi = {10.1103/RevModPhys.20.367},
  urldate = {2021-03-11}
}

@article{fixmanPolymerDynamicsBoson1966,
  title = {Polymer {{Dynamics}}: {{Boson Representation}} and {{Excluded}}-{{Volume Forces}}},
  shorttitle = {Polymer {{Dynamics}}},
  author = {Fixman, Marshall},
  year = {1966},
  month = aug,
  journal = {The Journal of Chemical Physics},
  volume = {45},
  number = {3},
  pages = {785--792},
  issn = {0021-9606},
  doi = {10.1063/1.1727682},
  urldate = {2025-08-01}
}

@book{fredricksonFieldTheoreticSimulationsSoft2023,
  title = {Field-{{Theoretic Simulations}} in {{Soft Matter}} and {{Quantum Fluids}}},
  author = {Fredrickson, Glenn H. and Delaney, Kris T. and Fredrickson, Glenn H. and Delaney, Kris T.},
  year = {2023},
  month = feb,
  series = {International {{Series}} of {{Monographs}} on {{Physics}}},
  publisher = {Oxford University Press},
  address = {Oxford, New York},
  isbn = {978-0-19-284748-5}
}

@article{galitskiSpinOrbitCoupling2013,
  title = {Spin--Orbit Coupling in Quantum Gases},
  author = {Galitski, Victor and Spielman, Ian B.},
  year = {2013},
  month = feb,
  journal = {Nature},
  volume = {494},
  number = {7435},
  pages = {49--54},
  publisher = {Nature Publishing Group},
  issn = {1476-4687},
  doi = {10.1038/nature11841},
  urldate = {2022-03-24},
  copyright = {2013 Nature Publishing Group, a division of Macmillan Publishers Limited. All Rights Reserved.},
  langid = {english}
}

@book{gerryIntroductoryQuantumOptics2004,
  title = {Introductory {{Quantum Optics}}},
  author = {Gerry, Christopher and Knight, Peter},
  year = {2004},
  publisher = {Cambridge University Press},
  address = {Cambridge},
  doi = {10.1017/CBO9780511791239},
  urldate = {2025-06-24},
  isbn = {978-0-521-52735-4}
}

@article{goldmanLightinducedGaugeFields2014,
  title = {Light-Induced Gauge Fields for Ultracold Atoms},
  author = {Goldman, N. and Juzeli{\=u}nas, G. and {\"O}hberg, P. and Spielman, I. B.},
  year = {2014},
  month = nov,
  journal = {Reports on Progress in Physics},
  volume = {77},
  number = {12},
  pages = {126401},
  publisher = {IOP Publishing},
  issn = {0034-4885},
  doi = {10.1088/0034-4885/77/12/126401},
  urldate = {2022-03-24},
  langid = {english}
}

@article{gopalakrishnanQuantumQuasicrystalsSpinOrbitCoupled2013,
  title = {Quantum {{Quasicrystals}} of {{Spin-Orbit-Coupled Dipolar Bosons}}},
  author = {Gopalakrishnan, Sarang and Martin, Ivar and Demler, Eugene A.},
  year = {2013},
  month = oct,
  journal = {Physical Review Letters},
  volume = {111},
  number = {18},
  pages = {185304},
  publisher = {American Physical Society},
  doi = {10.1103/PhysRevLett.111.185304},
  urldate = {2023-11-20}
}

@article{hayataComplexLangevinSimulation2015,
  title = {Complex {{Langevin}} Simulation of Quantum Vortices in a {{Bose-Einstein}} Condensate},
  author = {Hayata, Tomoya and Yamamoto, Arata},
  year = {2015},
  month = oct,
  journal = {Physical Review A},
  volume = {92},
  number = {4},
  pages = {043628},
  publisher = {American Physical Society},
  doi = {10.1103/PhysRevA.92.043628},
  urldate = {2024-06-01}
}

@article{heinenComplexLangevinApproach2022,
  title = {Complex {{Langevin}} Approach to Interacting {{Bose}} Gases},
  author = {Heinen, Philipp and Gasenzer, Thomas},
  year = {2022},
  month = dec,
  journal = {Physical Review A},
  volume = {106},
  number = {6},
  pages = {063308},
  publisher = {American Physical Society},
  doi = {10.1103/PhysRevA.106.063308},
  urldate = {2025-06-24}
}

@article{herdmanEntanglementAreaLaw2017,
  title = {Entanglement Area Law in Superfluid {{4He}}},
  author = {Herdman, C. M. and Roy, P.-N. and Melko, R. G. and Maestro, A. Del},
  year = {2017},
  month = jun,
  journal = {Nature Physics},
  volume = {13},
  number = {6},
  pages = {556--558},
  publisher = {Nature Publishing Group},
  issn = {1745-2481},
  doi = {10.1038/nphys4075},
  urldate = {2025-07-01},
  copyright = {2017 Springer Nature Limited},
  langid = {english}
}

@article{herdmanQuantumMonteCarlo2014,
  title = {Quantum {{Monte Carlo}} Measurement of the Chemical Potential of \$\{\}{\textasciicircum}\{4\}{\textbackslash}mathrm\{\vphantom\}{{He}}\vphantom\{\}\$},
  author = {Herdman, C. M. and Rommal, A. and Del Maestro, A.},
  year = {2014},
  month = jun,
  journal = {Physical Review B},
  volume = {89},
  number = {22},
  pages = {224502},
  publisher = {American Physical Society},
  doi = {10.1103/PhysRevB.89.224502},
  urldate = {2025-07-01}
}

@book{kamenevFieldTheoryNonEquilibrium2011,
  title = {Field {{Theory}} of {{Non-Equilibrium Systems}}},
  author = {Kamenev, Alex},
  year = {2011},
  month = sep,
  publisher = {Cambridge University Press},
  googlebooks = {CwlrUepnla4C},
  isbn = {978-1-139-50029-6},
  langid = {english}
}

@article{linSpinOrbitCoupled2011,
  title = {Spin--Orbit-Coupled {{Bose}}--{{Einstein}} Condensates},
  author = {Lin, Y.-J. and {Jim{\'e}nez-Garc{\'i}a}, K. and Spielman, I. B.},
  year = {2011},
  month = mar,
  journal = {Nature},
  volume = {471},
  number = {7336},
  pages = {83--86},
  publisher = {Nature Publishing Group},
  issn = {1476-4687},
  doi = {10.1038/nature09887},
  urldate = {2023-08-30},
  copyright = {2011 Springer Nature Limited},
  langid = {english}
}

@article{liStripePhaseSupersolid2017,
  title = {A Stripe Phase with Supersolid Properties in Spin--Orbit-Coupled {{Bose}}--{{Einstein}} Condensates},
  author = {Li, Jun-Ru and Lee, Jeongwon and Huang, Wujie and Burchesky, Sean and Shteynas, Boris and Top, Furkan {\c C}a{\u g}r{\i} and Jamison, Alan O. and Ketterle, Wolfgang},
  year = {2017},
  month = mar,
  journal = {Nature},
  volume = {543},
  number = {7643},
  pages = {91--94},
  publisher = {Nature Publishing Group},
  issn = {1476-4687},
  doi = {10.1038/nature21431},
  urldate = {2022-07-02},
  copyright = {2017 Macmillan Publishers Limited, part of Springer Nature. All rights reserved.},
  langid = {english}
}

@article{liTopologicalStatesLadderlike2013,
  title = {Topological States in a Ladder-like Optical Lattice Containing Ultracold Atoms in Higher Orbital Bands},
  author = {Li, Xiaopeng and Zhao, Erhai and Vincent Liu, W.},
  year = {2013},
  month = feb,
  journal = {Nature Communications},
  volume = {4},
  number = {1},
  pages = {1523},
  publisher = {Nature Publishing Group},
  issn = {2041-1723},
  doi = {10.1038/ncomms2523},
  urldate = {2025-07-01},
  copyright = {2013 Springer Nature Limited},
  langid = {english}
}

@article{mahmudFinitetemperatureStudyBosons2011,
  title = {Finite-Temperature Study of Bosons in a Two-Dimensional Optical Lattice},
  author = {Mahmud, K. W. and Duchon, E. N. and Kato, Y. and Kawashima, N. and Scalettar, R. T. and Trivedi, N.},
  year = {2011},
  month = aug,
  journal = {Physical Review B},
  volume = {84},
  number = {5},
  pages = {054302},
  publisher = {American Physical Society},
  doi = {10.1103/PhysRevB.84.054302},
  urldate = {2025-07-24}
}

@article{manchonNewPerspectivesRashba2015,
  title = {New Perspectives for {{Rashba}} Spin--Orbit Coupling},
  author = {Manchon, A. and Koo, H. C. and Nitta, J. and Frolov, S. M. and Duine, R. A.},
  year = {2015},
  month = sep,
  journal = {Nature Materials},
  volume = {14},
  number = {9},
  pages = {871--882},
  publisher = {Nature Publishing Group},
  issn = {1476-4660},
  doi = {10.1038/nmat4360},
  urldate = {2023-02-01},
  copyright = {2015 Nature Publishing Group, a division of Macmillan Publishers Limited. All Rights Reserved.},
  langid = {english}
}

@article{manCoherentStatesFormulation2014,
  title = {Coherent States Formulation of Polymer Field Theory},
  author = {Man, Xingkun and Delaney, Kris T. and Villet, Michael C. and Orland, Henri and Fredrickson, Glenn H.},
  year = {2014},
  month = jan,
  journal = {The Journal of Chemical Physics},
  volume = {140},
  number = {2},
  pages = {024905},
  publisher = {American Institute of Physics},
  issn = {0021-9606},
  doi = {10.1063/1.4860978},
  urldate = {2022-07-21}
}

@article{mcgarrigleEmergenceSpinMicroemulsion2023,
  title = {Emergence of a {{Spin Microemulsion}} in {{Spin-Orbit Coupled Bose-Einstein Condensates}}},
  author = {McGarrigle, Ethan C. and Delaney, Kris T. and Balents, Leon and Fredrickson, Glenn H.},
  year = {2023},
  month = oct,
  journal = {Physical Review Letters},
  volume = {131},
  number = {17},
  pages = {173403},
  publisher = {American Physical Society},
  doi = {10.1103/PhysRevLett.131.173403},
  urldate = {2023-10-26}
}

@article{mcgarrigleProjectedComplexLangevin2024,
  title = {Projected Complex {{Langevin}} Sampling Method for Bosons in the Canonical and Microcanonical Ensembles},
  author = {McGarrigle, Ethan C. and Ceniceros, Hector D. and Fredrickson, Glenn H.},
  year = {2024},
  month = dec,
  journal = {Physical Review E},
  volume = {110},
  number = {6},
  pages = {065308},
  publisher = {American Physical Society},
  doi = {10.1103/PhysRevE.110.065308},
  urldate = {2025-06-24}
}

@article{mcgarrigleThermodynamicStabilitySpin2024,
  title = {Thermodynamic Stability of a Spin Microemulsion in {{Rashba}} Spin-Orbit-Coupled Bosons},
  author = {McGarrigle, Ethan C. and Nodel, Ron and Delaney, Kris T. and Balents, Leon and Fredrickson, Glenn H.},
  year = {2024},
  month = oct,
  journal = {Physical Review A},
  volume = {110},
  number = {4},
  pages = {L041303},
  publisher = {American Physical Society},
  doi = {10.1103/PhysRevA.110.L041303},
  urldate = {2025-07-22}
}

@article{mukherjeeCrystallizationBosonicQuantum2022,
  title = {Crystallization of Bosonic Quantum {{Hall}} States in a Rotating Quantum Gas},
  author = {Mukherjee, Biswaroop and Shaffer, Airlia and Patel, Parth B. and Yan, Zhenjie and Wilson, Cedric C. and Cr{\'e}pel, Valentin and Fletcher, Richard J. and Zwierlein, Martin},
  year = {2022},
  month = jan,
  journal = {Nature},
  volume = {601},
  number = {7891},
  pages = {58--62},
  publisher = {Nature Publishing Group},
  issn = {1476-4687},
  doi = {10.1038/s41586-021-04170-2},
  urldate = {2025-07-01},
  copyright = {2022 The Author(s), under exclusive licence to Springer Nature Limited},
  langid = {english}
}

@book{negeleQuantumManyparticleSystems1988,
  title = {Quantum {{Many-particle Systems}}},
  author = {Negele, John W. and Orland, Henri},
  year = {1988},
  month = jan,
  publisher = {Basic Books},
  googlebooks = {EV8sAAAAYAAJ},
  isbn = {978-0-201-12593-1},
  langid = {english}
}

@article{norciaTwodimensionalSupersolidityDipolar2021,
  title = {Two-Dimensional Supersolidity in a Dipolar Quantum Gas},
  author = {Norcia, Matthew A. and Politi, Claudia and Klaus, Lauritz and Poli, Elena and Sohmen, Maximilian and Mark, Manfred J. and Bisset, Russell N. and Santos, Luis and Ferlaino, Francesca},
  year = {2021},
  month = aug,
  journal = {Nature},
  volume = {596},
  number = {7872},
  pages = {357--361},
  issn = {0028-0836, 1476-4687},
  doi = {10.1038/s41586-021-03725-7},
  urldate = {2022-08-05},
  langid = {english}
}

@article{pelitiPathIntegralApproach1985,
  title = {Path Integral Approach to Birth-Death Processes on a Lattice},
  author = {Peliti, L.},
  year = {1985},
  month = sep,
  journal = {Journal de Physique},
  volume = {46},
  number = {9},
  pages = {1469--1483},
  publisher = {Soci{\'e}t{\'e} Fran{\c c}aise de Physique},
  issn = {0302-0738, 2777-3396},
  doi = {10.1051/jphys:019850046090146900},
  urldate = {2025-08-01},
  langid = {english}
}

@book{pitaevskiiBoseEinsteinCondensationSuperfluidity2016,
  title = {Bose-{{Einstein Condensation}} and {{Superfluidity}}},
  author = {Pitaevskii, Lev and Stringari, Sandro},
  year = {2016},
  month = jan,
  publisher = {Oxford University Press},
  googlebooks = {yHByCwAAQBAJ},
  isbn = {978-0-19-107668-8},
  langid = {english}
}

@article{polletRecentDevelopmentsQuantum2012,
  title = {Recent Developments in Quantum {{Monte Carlo}} Simulations with Applications for Cold Gases},
  author = {Pollet, Lode},
  year = {2012},
  month = aug,
  journal = {Reports on Progress in Physics},
  volume = {75},
  number = {9},
  pages = {094501},
  publisher = {IOP Publishing},
  issn = {0034-4885},
  doi = {10.1088/0034-4885/75/9/094501},
  urldate = {2025-07-28},
  langid = {english}
}

@article{rubensteinFinitetemperatureAuxiliaryfieldQuantum2012,
  title = {Finite-Temperature Auxiliary-Field Quantum {{Monte Carlo}} Technique for {{Bose-Fermi}} Mixtures},
  author = {Rubenstein, Brenda M. and Zhang, Shiwei and Reichman, David R.},
  year = {2012},
  month = nov,
  journal = {Physical Review A},
  volume = {86},
  number = {5},
  pages = {053606},
  publisher = {American Physical Society},
  doi = {10.1103/PhysRevA.86.053606},
  urldate = {2023-10-04}
}

@book{sachdevQuantumPhaseTransitions2001,
  title = {Quantum {{Phase Transitions}}},
  author = {Sachdev, Subir},
  year = {2001},
  month = apr,
  publisher = {Cambridge University Press},
  googlebooks = {Ih\_E05N5TZQC},
  isbn = {978-0-521-00454-1},
  langid = {english}
}

@article{semeghiniProbingTopologicalSpin2021,
  title = {Probing Topological Spin Liquids on a Programmable Quantum Simulator},
  author = {Semeghini, G. and Levine, H. and Keesling, A. and Ebadi, S. and Wang, T. T. and Bluvstein, D. and Verresen, R. and Pichler, H. and Kalinowski, M. and Samajdar, R. and Omran, A. and Sachdev, S. and Vishwanath, A. and Greiner, M. and Vuleti{\'c}, V. and Lukin, M. D.},
  year = {2021},
  month = dec,
  journal = {Science},
  volume = {374},
  number = {6572},
  pages = {1242--1247},
  publisher = {American Association for the Advancement of Science},
  doi = {10.1126/science.abi8794},
  urldate = {2025-07-01}
}

@article{shenFiniteTemperatureAuxiliary2020,
  title = {Finite Temperature Auxiliary Field Quantum {{Monte Carlo}} in the Canonical Ensemble},
  author = {Shen, Tong and Liu, Yuan and Yu, Yang and Rubenstein, Brenda M.},
  year = {2020},
  month = nov,
  journal = {The Journal of Chemical Physics},
  volume = {153},
  number = {20},
  pages = {204108},
  issn = {0021-9606},
  doi = {10.1063/5.0026606},
  urldate = {2023-10-03}
}

@article{shenStableRecursiveAuxiliary2023,
  title = {Stable Recursive Auxiliary Field Quantum {{Monte Carlo}} Algorithm in the Canonical Ensemble: {{Applications}} to Thermometry and the {{Hubbard}} Model},
  shorttitle = {Stable Recursive Auxiliary Field Quantum {{Monte Carlo}} Algorithm in the Canonical Ensemble},
  author = {Shen, Tong and Barghathi, Hatem and Yu, Jiangyong and Del Maestro, Adrian and Rubenstein, Brenda M.},
  year = {2023},
  month = may,
  journal = {Physical Review E},
  volume = {107},
  number = {5},
  pages = {055302},
  publisher = {American Physical Society},
  doi = {10.1103/PhysRevE.107.055302},
  urldate = {2024-07-30}
}

@article{shumwayPathIntegralMonte2000,
  title = {Path Integral {{Monte Carlo}} Simulations for Fermion Systems : {{Pairing}} in the Electron-Hole Plasma},
  shorttitle = {Path Integral {{Monte Carlo}} Simulations for Fermion Systems},
  author = {Shumway, J. and Ceperley, D. M.},
  year = {2000},
  month = mar,
  journal = {Le Journal de Physique IV},
  volume = {10},
  number = {PR5},
  pages = {Pr5-3-Pr5-16},
  issn = {1155-4339},
  doi = {10.1051/jp4:2000501},
  urldate = {2023-04-05},
  langid = {english}
}

@article{valdes-curielTopologicalFeaturesLattice2021,
  title = {Topological Features without a Lattice in {{Rashba}} Spin-Orbit Coupled Atoms},
  author = {{Vald{\'e}s-Curiel}, A. and Trypogeorgos, D. and Liang, Q.-Y. and Anderson, R. P. and Spielman, I. B.},
  year = {2021},
  month = jan,
  journal = {Nature Communications},
  volume = {12},
  number = {1},
  pages = {593},
  publisher = {Nature Publishing Group},
  issn = {2041-1723},
  doi = {10.1038/s41467-020-20762-4},
  urldate = {2023-03-03},
  copyright = {2021 This is a U.S. government work and not under copyright protection in the U.S.; foreign copyright protection may apply},
  langid = {english}
}

@article{wilsonMeronGroundState2013,
  title = {Meron {{Ground State}} of {{Rashba Spin-Orbit-Coupled Dipolar Bosons}}},
  author = {Wilson, Ryan M. and Anderson, Brandon M. and Clark, Charles W.},
  year = {2013},
  month = oct,
  journal = {Physical Review Letters},
  volume = {111},
  number = {18},
  pages = {185303},
  publisher = {American Physical Society},
  doi = {10.1103/PhysRevLett.111.185303},
  urldate = {2025-07-24}
}

@article{zillichExtrapolatedHighorderPropagators2010,
  title = {Extrapolated High-Order Propagators for Path Integral {{Monte Carlo}} Simulations},
  author = {Zillich, Robert E. and Mayrhofer, Johannes M. and Chin, Siu A.},
  year = {2010},
  month = jan,
  journal = {The Journal of Chemical Physics},
  volume = {132},
  number = {4},
  pages = {044103},
  issn = {0021-9606},
  doi = {10.1063/1.3297888},
  urldate = {2025-07-02}
}

@misc{si, title={See supplementary material for additional details on the transformation of coherent-state wavefunctions under quadratic propagators, the application of this approach to Rashba spin-orbit-coupled bosons, and the estimation of observables in coherent state complex Langevin simulations.}}


\end{document}